\documentclass[aps,pra,reprint,superscriptaddress,longbibliography]{revtex4-2}
\usepackage[utf8]{inputenc}
\usepackage[english]{babel}
\usepackage[T1]{fontenc}
\usepackage{amssymb}
\usepackage{hyperref}
\usepackage{float}
\usepackage{enumitem}
\usepackage{tikz}
\usetikzlibrary{quantikz}
\usepackage{bbold}
\usepackage{natbib}
\usepackage{cancel}
\newtheorem{theorem}{Theorem}
\usepackage[normalem]{ulem}
\begin{document}
\definecolor{colorx}{RGB}{137,107,186}
\definecolor{colory}{RGB}{53,97,14}
\definecolor{colorz}{RGB}{233,169,41}  
\renewcommand{\figurename}{FIG.}

\title{Digital-analog quantum computation with arbitrary two-body Hamiltonians}

\author{Mikel Garcia-de-Andoin}\email{mikel.garciadeandoin@\{tecnalia.com/ehu.eus\}}\affiliation{Department of Physical Chemistry, University of the Basque Country UPV/EHU, Apartado 644, 48940 Leioa, Spain}\affiliation{TECNALIA, Basque Research and Technology Alliance (BRTA), Astondo Bidea Edificio 700, 48160 Derio, Spain}\affiliation{EHU Quantum Center, University of the Basque Country UPV/EHU, Barrio Sarriena s/n, 48940 Leioa, Spain}
\author{\'Alvaro Saiz}\email{asaiz@us.es}\affiliation{Departamento de Física Aplicada III, Universidad de Sevilla, Camino Descubrimientos s/n, 41092 Sevilla, Spain}
\author{Pedro P\'erez-Fern\'andez}\affiliation{Departamento de Física Aplicada III, Universidad de Sevilla, Camino Descubrimientos s/n, 41092 Sevilla, Spain}\affiliation{Instituto Carlos I de F\'isica Te\'orica y Computacional, Universidad de Granada, Av. de Fuente Nueva s/n, 18071 Granada, Spain}
\author{Lucas Lamata}\affiliation{Departamento de Física Atómica, Molecular y Nuclear, Universidad de Sevilla, Av. de la Reina Mercedes s/n, 41012 Sevilla, Spain}\affiliation{Instituto Carlos I de F\'isica Te\'orica y Computacional, Universidad de Granada, Av. de Fuente Nueva s/n, 18071 Granada, Spain}
\author{Izaskun Oregi}\affiliation{TECNALIA, Basque Research and Technology Alliance (BRTA), Astondo Bidea Edificio 700, 48160 Derio, Spain}
\author{Mikel Sanz}\affiliation{Department of Physical Chemistry, University of the Basque Country UPV/EHU, Apartado 644, 48940 Leioa, Spain}\affiliation{EHU Quantum Center, University of the Basque Country UPV/EHU, Barrio Sarriena s/n, 48940 Leioa, Spain}\affiliation{IKERBASQUE, Basque Foundation for Science, Plaza Euskadi 5, 48009 Bilbao, Spain}\affiliation{Basque Center for Applied Mathematics (BCAM), Alameda Mazarredo 14, 48009 Bilbao, Spain}

\date{\today}

\begin{abstract}
Digital-analog quantum computing is a computational paradigm which employs an analog Hamiltonian resource together with single-qubit gates to reach universality. Here, we design a new scheme which employs an arbitrary two-body source Hamiltonian, extending the experimental applicability of this computational paradigm to most quantum platforms. We show that the simulation of an arbitrary two-body target Hamiltonian of $n$ qubits requires $\mathcal{O}(n^2)$ analog blocks with guaranteed positive times, providing a polynomial advantage compared to the previous scheme. Additionally, we propose a classical strategy which combines a Bayesian optimization with a gradient descent method, improving the performance by $\sim55\%$ for small systems measured in the Frobenius norm.
\end{abstract}

\maketitle

\section{\label{sec:intro}Introduction}

When quantum computing was originally proposed~\cite{Benioff1980, Feynman}, it was envisioned as a way of simulating the dynamics of a quantum system employing another controllable system. This set the foundations of what we now call analog quantum computing (AQC)~\cite{Arnab2008}. A different approach was introduced when Deutsch proposed the concept of a quantum gate~\cite{Deutsch1995}, which finally led to the digital quantum computing (DQC) paradigm. 

One of the main advantages of AQC is the robustness of the simulation. Quantum control techniques have been developed in the last decades, providing a high fidelity and further protecting the dynamics against different sources of errors~\cite{Werschnik2007, Koch2022}. Despite their robustness, AQC is strongly limited by the dynamics of the system, making it difficult to implement most dynamics of interest. In contrast, DQC is performed through the sequential application of quantum gates in a discrete manner, mimicking classical computations. It is proven that any unitary can be decomposed with arbitrary precision in terms of single-qubit gates (SQG) and at least one entangling two-qubit gate (TQG)~\cite{NielsenChuang2010}. One of the main features that DQC provides is the possibility of applying quantum error correction (QEC) techniques~\cite{Kitaev1997}. In the current Noisy Intermediate-Scale Quantum (NISQ) era~\cite{Preskill2018NISQ}, the qubits and the gates available are noisy and prone to errors. Thus, the only hope of reaching fault tolerant quantum computing is through the application of sophisticated QEC techniques~\cite{Bultrini2022} once we fulfill the requirements for the quantum threshold theorem~\cite{aharonov1996, knill1996, Kitaev1997}.

The digital-analog quantum computing (DAQC) paradigm was proposed as a way of combining the robustness of AQC with the versatility of DQC, ~\cite{Lamata2018, Adrian2020DAQC}. The main idea behind DAQC is employing the natural interaction Hamiltonian of a system as an entanglement resource. By alternating the evolution under this Hamiltonian (analog blocks) and the application of SQGs (digital blocks), one can simulate an arbitrary target Hamiltonian. Here, we can distinguish two kinds of approaches. If the interaction Hamiltonian is turned off during the application of the digital blocks, we call this approach stepwise-DAQC (sDAQC) circuit. Otherwise, if for practical purpose the system Hamiltonian is always on, and the SQGs are performed on top of this dynamics, we call this approach banged-DAQC (bDAQC) circuits. Interestingly, although this introduces a systematic error, this scales better than main error sources found in quantum computers~\cite{garcia2022noise}. It has already been experimentally proven that DAQC is a suitable paradigm for the NISQ era, for instance, in the implementation of a variational quantum algorithm in a system with up to 61 qubits~\cite{WeiPan202261qubitDAQC}.

In order to enhance the range of quantum platforms suitable for DAQC, we must extend the techniques to arbitrary resource Hamiltonians. Previously, the resource Hamiltonian was the aforementioned Ising Hamiltonian. Additionally, the construction of an arbitrary target Hamiltonian was performed through a two-step procedure. First, by transforming the source Hamiltonian into an adequate ZZ Hamiltonian~\cite{Tasio2021DAQCcrossResonance}. Then, by employing sequences of SQGs to build an arbitrary Hamiltonian~\cite{Adrian2020DAQC}. However, the question of systematically performing this transformation in a single step was still open.

In this article, we extend the DAQC protocol to approximate the evolution under an arbitrary target two-body Hamiltonian by evolving under another arbitrary two-body resource Hamiltonian up to a certain Trotter error. By means of the Trotter-Suzuki formula, we argue that by repeating this sequence $n_T$ times one can reduce the error exponentially with $n_T$. The tools we develop in this article allow for a practical realization of DAQC schedules in faulty hardware, in which spurious couplings prevent us from approximating the system as an Ising Hamiltonian. We also solve the problem of the negative analog blocks times that limited the implementability of previous protocols. Additionally, we introduce a classical optimization technique to find optimal angles for the SQGs. Taking into account the depth limitations of quantum circuits in the NISQ era, our objective is to maximize the fidelity of the circuit employing a fixed amount of digital-analog blocks. We show how it is possible to employ DAQC schedules with a low number of digital-analog blocks that achieves fidelities compared to a systematic approach with higher count of blocks.

The rest of the article  is organized in the following manner. In Sec.~\ref{sec:warmup}, we review the previous protocol employing $ZZ$ Hamiltonians. In Sec.~\ref{sec:Extension}, we present the new protocol which extends it to an arbitrary two-body source Hamiltonian, and discuss the error scaling. Then, in Sec.~\ref{sec:optimizing}, we introduce an optimization technique for approximating arbitrary dynamics employing a fixed number of blocks, and illustrate the technique for a particular problem. Finally, in Sec.~\ref{sec:Conclusions} we conclude with some final remarks.

\section{\label{sec:warmup}Warm-up: DAQC protocol for $ZZ$ Hamiltonians}

As a warm up, let us review the previous protocol for simulating the dynamics during a time $T$ of a target $ZZ$ all-to-all (ATA) Hamiltonian
\begin{equation}
    H_\text{T,ZZ}=\sum_{i<j}^n g_{i,j}\sigma_i^z\sigma_j^z,
\end{equation}
by employing a source $ZZ$ ATA Hamiltonian
\begin{equation}
    H_\text{S,ZZ}=\sum_{i<j}^n h_{i,j}\sigma_i^z\sigma_j^z.
\end{equation}

\begin{figure}[t]  
    \centering
    \resizebox{\linewidth}{!}{
    \begin{quantikz}[row sep={0.7cm,between origins},column sep=0.25cm]
        & \gate[style={fill=colorx!20}]{X} & \gate[wires=5]{t_{1,2}} & \gate[style={fill=colorx!20}]{\cancel{X}} & \gate[style={fill=colorx!20}]{\cancel{X}} & \gate[wires=5]{t_{1,3}} & \gate[style={fill=colorx!20}]{X} & \ \phantom{\ldots}\ \qw & \qw              & \gate[wires=5]{t_{4,5}} & \qw              & \qw \\
        & \gate[style={fill=colorx!20}]{X} &                                         & \gate[style={fill=colorx!20}]{X} & \qw              &                                         & \qw              & \ \phantom{\ldots}\ \qw & \qw              &                                         & \qw              & \qw \\
        & \qw              &                                         & \qw              & \gate[style={fill=colorx!20}]{X} &                                         & \gate[style={fill=colorx!20}]{X} & \ \ldots\ \qw           & \qw              &                                         & \qw              & \qw \\ 
        & \qw              &                                         & \qw              & \qw              &                                         & \qw              & \ \phantom{\ldots}\ \qw & \gate[style={fill=colorx!20}]{X} &                                         & \gate[style={fill=colorx!20}]{X} & \qw \\
        & \qw              &                                         & \qw              & \qw              &                                         & \qw              & \ \phantom{\ldots}\ \qw & \gate[style={fill=colorx!20}]{X} &                                         & \gate[style={fill=colorx!20}]{X} & \qw \\
    \end{quantikz}}
    \caption{sDAQC circuit for the Ising $ZZ$ ATA Hamiltonian for 5 qubits. For simulating an arbitrary evolution, we sandwich several analog blocks with a couple of $X$ gates applied to all combinations of two qubits. The blocks labeled with $t_{i,j}$ represent the unitary evolution under the source Hamiltonian for the time $t_{i,j}$, this is $e^{-it_{i,j}H_\text{S}}$.}
    \label{fig:zz_protocol}
\end{figure}
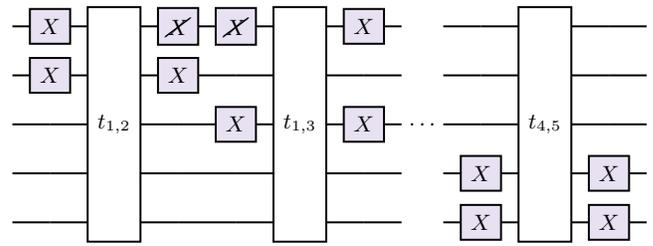

For achieving this, in Refs.~\cite{Adrian2020DAQC,Galicia2020EnhancedConnect} the authors proposed a universal protocol, pictorially shown in Fig.~\ref{fig:zz_protocol}. It consists in sandwiching each analog block with two $X$ gates, applied to a different pair of qubits each time. Effectively, this changes the sign of all couplings in which only one of the qubits is selected. Noticing that all terms of the Hamiltonian commute with each other, we have that the Trotter formula is exact, so we can write\small
\begin{eqnarray}\label{eq:zz_protocol}
    U_\text{T}&=&e^{-iTH_\text{T}}\nonumber\\
    &=&\prod_{i<j}\exp\left(-it_{i,j}\sum_{\ell<m}(-1)^{\delta_{i\ell}+\delta_{im}+\delta_{j\ell}+\delta_{jm}}h_{\ell,m}\sigma^z_\ell\sigma^z_m\right),\nonumber\\
\end{eqnarray}
\normalsize where $\delta_{ij}$ is the Kronecker delta and $t_{i,j}$ is the analog time for the corresponding analog block. Note that throughout this work, we are considering $\hbar=1$. We can rewrite the equation more conveniently as a linear system of equations
\begin{equation}\label{eq:systemEqs}
    M\vec{t}=T \overrightarrow{g/h},
\end{equation}
where the elements of the matrix $M_{(i,j),(\ell,m)}=(-1)^{\delta_{i\ell}+\delta_{im}+\delta_{j\ell}+\delta_{jm}}$ represents the effective signs of the couplings between qubits $(i,j)$ in every analog block $(\ell,m)$, $\vec{t}$ is the vector of times of each analog block, and $(g/h)_{(i,j)}$ is a vector of the proportion between the target and the source coupling strengths between the qubits $(i,j)$. If there is a missing coupling in both source and target Hamiltonians, $g_{i,j}=h_{i,j}=0$, then we remove the corresponding element from the vector $\overrightarrow{g/h}$ and the corresponding row in $M$. If the coupling is missing but the target coupling is non-zero, then the Hamiltonian cannot be simulated. It can be proven that the matrix $M$ is non-singular for all number of qubits except 4, so we can obtain an exact simulation of the desired dynamics employing this schedule.

\section{\label{sec:Extension}Extension of DAQC to arbitrary two-body Hamiltonians}

The proof that almost any entangling two-body Hamiltonian, together with SQGs, can be employed to simulate the dynamics of another two-body Hamiltonian was shown in Ref.~\cite{Dodd2002UnivQC}, but no constructive method was provided. From now on, we will refer to the source Hamiltonian
\begin{equation}
    H_\text{S}=\sum_{i<j}^n\sum_{\mu,\nu\in\{x,y,z\}}h_{i,j}^{\mu,\nu}\sigma^\mu_i\sigma^\nu_j,
\end{equation}
and the target Hamiltonian as
\begin{equation}
    H_\text{T}=\sum_{i<j}^n\sum_{\mu,\nu\in\{x,y,z\}}g_{i,j}^{\mu,\sigma}\sigma^\mu_i\sigma^\nu_j,
\end{equation}
where $\sigma^\mu_i$ is the Pauli operator $\mu$ acting on qubit $i$. Then, the objective is to obtain a circuit to simulate the evolution of $H_\text{T}$ for a time $T$, $U=e^{-iTH_\text{T}}$. The first step of the proof is to note that it is possible to decouple a pair of qubits from the rest in an $n$-qubit system. Then, by employing a 36-step digital-analog protocol, it can be proven that any two-qubit interaction can be simulated. The proof for universality can be obtained by extending this to every coupling in the target Hamiltonian, and repeating the circuit $n_\text{T}$ times for simulating a time $\Delta=T/n_\text{T}$ in each Trotter step. However, this protocol gets convoluted as the number of qubits in the system, $n$, increases. In general, the circuit requires $\mathcal{O}(n^3n_\text{T})$ analog blocks, with an error of $\varepsilon\sim\mathcal{O}(n^2 T^2/n_T)$. As it stands, the question of obtaining a more efficient protocol is still open.

\begin{figure}[ht]
    \resizebox{\linewidth}{!}{
    \begin{tikzpicture}
    \node (a) at (0,5) {
        \begin{quantikz}[row sep={0.7cm,between origins},column sep=0.8cm]
        & \gate{\phantom{j}}\gategroup[5,steps=1,style={rounded corners}]{1,2} & \gate{\phantom{j}}\gategroup[5,steps=1,style={rounded corners}]{1,3} & \ \phantom{\ldots}\ \qw  & \hphantomgate{j}\gategroup[5,steps=1,style={rounded corners}]{$i,j$} & \ \phantom{\ldots}\ \qw  & \qw\gategroup[5,steps=1,style={rounded corners}]{4,5} & \qw \\
        & \gate{} & \qw & \ \phantom{\ldots}\ \qw & \qw & \ \phantom{\ldots}\ \qw & \qw & \qw \\
        & \qw & \gate{} & \ \ldots\ \qw & \qw & \ \ldots\ \qw & \qw & \qw \\
        & \qw & \qw & \ \phantom{\ldots}\ \qw & \qw & \ \phantom{\ldots}\ \qw & \gate{\phantom{j}} & \qw \\
        & \qw & \qw & \ \phantom{\ldots}\ \qw & \qw & \ \phantom{\ldots}\ \qw & \gate{} & \qw
        \end{quantikz}};
    \node (b) at (0,0) {
        \begin{quantikz}[row sep={0.7cm,between origins},column sep=0.25cm]
        & \gate[style={fill=colorx!20}]{X} & \gate[wires=5]{t_{1,3}^{x,x}} & \gate[style={fill=colorx!20}]{\cancel{X}} & \gate[style={fill=colorx!20}]{\cancel{X}} & \gate[wires=5]{t_{1,3}^{x,y}} & \gate[style={fill=colorx!20}]{X} & \ \phantom{\ldots}\ \qw &  \gate[style={fill=colorz!20}]{Z} & \gate[wires=5]{t_{1,3}^{z,z}} & \gate[style={fill=colorz!20}]{Z} & \qw \\
        & \qw & \qw & \qw & \qw & \qw & \qw & \ \phantom{\ldots}\ \qw & \qw & \qw & \qw & \qw \\
        & \gate[style={fill=colorx!20}]{X} & \qw & \gate[style={fill=colorx!20}]{X} & \gate[style={fill=colory!20}]{Y} & \qw & \gate[style={fill=colory!20}]{Y} & \ \ldots\ \qw &  \gate[style={fill=colorz!20}]{Z} & \qw & \gate[style={fill=colorz!20}]{Z} & \qw \\
        & \qw & \qw & \qw & \qw & \qw & \qw & \ \phantom{\ldots}\ \qw & \qw & \qw & \qw & \qw \\
        & \qw & \qw & \qw & \qw & \qw & \qw & \ \phantom{\ldots}\ \qw & \qw & \qw & \qw & \qw \\
        \end{quantikz}};
    \draw[thick] (b.north west) -- (-2.6,2.8);
    \draw[thick] (b.north east) -- (-1.4,2.8);
    \end{tikzpicture}}
    \caption{sDAQC circuit for an arbitrary ATA Hamiltonian for 5 qubits. The protocol for simulating any target Hamiltonian can be divided into two steps. In the first step (top), we select each of the $n(n-1)/2$ pair of qubits in our system. In the second step (bottom), for each pair of qubits, we apply all the 9 possible combinations of the Pauli gates. The number of analog blocks for this protocol scales quadraticaly with the number of qubits, $\mathcal{O}(n^2)$.}
    \label{fig:arbitrary_protocol}
\end{figure}
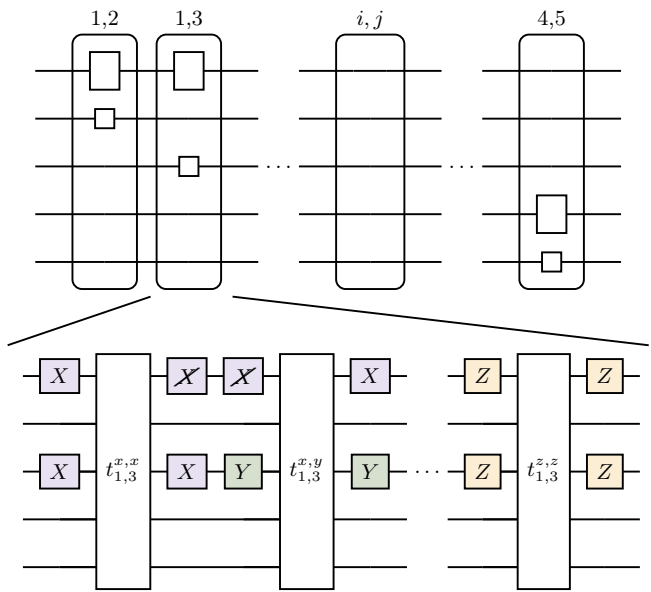

For the general case, we can extend the ideas reviewed in Sec.~\ref{sec:warmup} and extend it to arbitrary two-body Hamiltonians. The first step for our protocol is to select each pair of qubits $\{i,j\}$ of our system. Now, instead of just applying an $X$ gate to both of them, we will apply all 9 possible choices of pairs of gates $\{XX,XY,XZ,YX,YY,YZ,ZX,ZY,ZZ\}$. This is illustrated for a simple example in Fig.~\ref{fig:arbitrary_protocol}. Applying this to every pair of qubits, we effectively change the sign of some of the couplings, generating a non-singular system of $9n(n-1)/2$ equations. As the number of equations coincides with the number of variables and the number of parameters to define an arbitrary two-body Hamiltonian, this protocol is optimal in the number of digital-analog blocks. Unfortunately, in the general case the Hamiltonian does not commute with itself. This means that unlike the warm-up case, the Trotter formula is not exact, and thus, if we want to achieve an arbitrarily small error, we need to employ more Trotter steps. As a quick sketch of the proof, we write the problem as a system of equations similar to the one in Eq.~\ref{eq:systemEqs},
\begin{equation}\label{eq:systemEqsNew}
    M(n)\vec{t}=T\overrightarrow{g/h}.
\end{equation}
The matrix for a $n$-qubit system can be constructed recursively as a block matrix,
\begin{equation}
    M(n)=\begin{pmatrix}
    A(n) & P(n)\\
    Q(n) & M(n-1)
    \end{pmatrix},
\end{equation}
where the blocks $P(n)$, $Q(n)$ and $P(n)$ can be constructed systematically by taking into account the change of signs of the effective Hamiltonian terms after sandwiching them by Pauli gates. Then, by using the properties of this matrix and the definition of the formal determinant, we can prove that it is non-singular. Further details for the proof of the universality of the protocol are given in Appendix~\ref{Apx:universality}. With this result, we can then employ the same results as in the original work by Suzuki~\cite{Suzuki1976Trotter} to argue that an arbitrarily small error can be attained. 

\subsection{Analysis of the errors}

In order to obtain a bound for the maximum error of this protocol, we can resort to the original error analysis of the Suzuki-Trotter formula~\cite{Suzuki1976Trotter}. Since we have proven that the sum of the effective Hamiltonians in each block is exactly the target Hamiltonian, we can employ the formula for the $(n_\text{T},1)$ approximant
\begin{eqnarray}\label{eq:approximant}
    \varepsilon&=&\left\lVert U_\text{T}-U_\text{S}\right\rVert
    =\left\lVert e^{-iTH_\text{T}}-\left(\prod_ke^{-i\frac{t_k}{n_T}H_\text{S}^{(k)}}\right)^{n_\text{T}}\right\rVert\nonumber\\
    &\leq& \frac{2}{n_\text{T}}\left(\sum_k\lVert t_kH_\text{S}^{(k)}\rVert\right)^2e^{\frac{n_\text{T}+2}{n_\text{T}}\sum_k\lVert t_kH_\text{S}^{(k)}\rVert},
\end{eqnarray}
where $H_\text{S}^{(k)}$ is the effective Hamiltonian in the $k$-th analog block and  $\lVert \cdot \rVert$ is the Frobenius norm defined as $\lVert A\rVert=\sqrt{AA^\dagger}$. We will employ this norm for matrices throughout the text.

Since in each block only the sign of some Pauli string terms in the Hamiltonian changes, the norm is the same for all blocks $\lVert H_\text{S}^{(k)}\rVert=\lVert H_\text{S}\rVert$. With this, we have that the error is bounded by the sum of the times of the analog blocks. If we assume that we have a correct protocol in which all the time of the analog blocks are positive, we can rewrite
\begin{equation}
    \varepsilon \leq \frac{2}{n_\text{T}}t_\text{A}^2\lVert H_\text{S}\rVert^2e^{\frac{n_\text{T}+2}{n_\text{T}}t_\text{A}\lVert H_\text{S}\rVert},
\end{equation}
where $t_\text{A}=\sum_kt_k$. 

The total analog time will be lower bounded by the norms of $H_\text{S}$ and $H_\text{T}$ and the time $T$ of the simulation, $t_\text{A}\geq T \lVert H_\text{T}\rVert/\lVert H_\text{S}\rVert$. This corresponds to a situation in which both Hamiltonians are proportional to each other, $H_\text{S}\sim H_\text{T}$. Let us now study a general case. Assume an optimal protocol in which the total time is minimized over all possible protocols. In this case, $t_\text{A}$ is upper bounded by the weakest source coupling and the strongest target coupling as 
\begin{equation}
    t_\text{A}\leq C(n)\ T\ \frac{\max_{(i,j,\mu,\nu)}\lvert g_{i,j}^{\mu,\nu}\rvert}{\min_{(i,j,\mu,\nu)}\lvert h_{i,j}^{\mu,\nu}\rvert},
\end{equation}
where $C(n)\geq 1$ is a constant depending both on the protocol and the system size. In the case a coupling $h_{i,j}^{\mu,\nu}$ is 0 we take it out the corresponding element from both $\vec{g}$ and $\vec{h}$ and the corresponding row from the matrix $M$. In general, the constant $C(n)$ heavily depends on the protocol, but it does not monotonically grow with the number of qubits. For example, we find that for the original protocol in Ref.~\cite{Adrian2020DAQC} is upper bounded by $C(n)=n(n-1)/(n-4)(n-5)\leq15$ when $n>5$ for a balanced solution ($\vec{t}\sim(1,\dots,1)$). Similarly, the protocol in Ref.~\cite{Galicia2020EnhancedConnect} for a nearest-neighbour Hamiltonian has $C(n)\leq3/2$ for any system size.

Employing this result, we see that we can make the error arbitrarily small by increasing the number of Trotter steps of the protocol. However, this would only work for sDAQC circuits. When working on the bDAQC paradigm increasing the number of Trotter steps would increase the bang error in the protocol, as the time to apply the SQGs remains the same. This means that there is a point at which increasing the number of Trotter steps actually reduces the fidelity of the circuit. However, this analysis should be performed case by case, as it heavily depends on the problem and the system characteristics. As a rule of thumb, the time for the shortest analog block in the circuit should be at least two orders of magnitude above the time for applying a SQG, $\text{min}(t_k)\gtrsim 10^2t_\text{SQG}$.

\subsection{A note about negative times}

When computing the times of the analog blocks in Eq.~\ref{eq:systemEqs}, one can obtain a solution comprising some negative times. Simulating the evolution over a negative time would require to completely flip the sign of all the terms in the corresponding $H_\text{S}^{(k)}$. However, this cannot be done in general. For instance, one can straightforwardly prove that one can not do this for the 3 qubit all-to-all connected system by exhausting all possible combinations. As a suggestion for solving this problem, it was originally proposed to add one extra analog block, without it been sandwiched by any SQG. However, this only solves the problem in some particular cases. Approximate solutions to similar problems have been proposed~\cite{Pascal2023}, but the question of finding a systematic solution was still open.

Here we propose a method for constructing DAQC protocols in which the times are all positive. This method is highly inefficient in single qubit gate counts, but useful for proving the existence of such solution. The problem is exactly the same as in Eq.~\ref{eq:systemEqsNew} but, instead of a square matrix $M(n)$, we will employ all combinations of Pauli gates plus the identity $\{\mathbb{1},X,Y,Z\}$ to construct a matrix with $4^n$ different columns $M_i$. Then, we map the problem to a non-negative least-squares (NNLS) problem, for which we then employ the Algorithm NNLS to obtain a positive solution~\cite{Lawson1974}. However, we first need to prove that a positive solution exists. Here we roughly sketch the proof for this claim. Firstly, we note that the columns $M_i$ corresponds to the vertices of a polytope. Secondly, we prove that there is a strictly positive solution for the homogeneous system $\overrightarrow{g/h}=\vec{0}$, with $\vec{t}=4^{-n}\vec{1}$. Lastly, we build a hypersphere centered in $\vec{0}$ with small radius with a strictly positive vector $\vec{s}=4^{-n}\vec{1}+\vec{\varepsilon}$, from which we can reach to any possible problem $\overrightarrow{g/h}$ by scaling with a positive number, $\overrightarrow{g/h}=\lambda\vec{s}$ for $\lambda>0$. Even though we are using an exponential number of blocks for the proof, numerically we observe that the solution contains only $\sim 9n(n-1)/2$ nonzero elements. The full proof is provided in Appendix~\ref{Apx:PositiveTimes}.

\section{\label{sec:optimizing}Classical optimization of the DAQC schedule}
The protocol discussed in the previous section is a systematic method to obtain an arbitrary target Hamiltonian using another arbitrary Hamiltonian as a source. Its implementation involves a number of digital-analog blocks that grows quadratically with the number of qubits. With the current limitations of NISQ devices, long circuits can accumulate large experimental errors, so it becomes necessary to find a trade-off between the accuracy of the theoretical approximation and the required experimental resources. 

With the goal of reducing the number of digital-analog blocks, we propose a classical optimization strategy to find a set of SQGs sandwiching $K$ analog blocks such that the digital-analog schedule is as close as possible to the ideal evolution. This way, we propose an optimization problem where the parameters to be optimized are the times of the analog blocks $t_k$ and the parameters of an arbitrary SQG 
\begin{equation}\label{eq:ArbitraryRotation}
    R(\theta,\phi,\lambda) = 
    \begin{pmatrix}
    \cos(\frac{\theta}{2})           & -e^{i\lambda}\sin(\frac{\theta}{2}) \\ \\
    e^{i\phi} \sin(\frac{\theta}{2}) & e^{i(\lambda + \phi)} \cos(\frac{\theta}{2})
    \end{pmatrix},
\end{equation}
where $\{\theta, \phi, \lambda\}\in[0,2\pi)$ are the rotation angles of the SQG. The cost function we minimize is the Frobenius distance between the target evolution $U_\text{T}$ and the circuit with the optimized parameters $U_\text{C}$, similar to the calculation in Eq.~\ref{eq:approximant}. By employing the Frobenius distance between the unitaries as a proxy for the fidelity, we can test the approach in general, without imposing any assumptions about the initial state of the system or without expensive Haar integral calculations.

For simulating the circuits, we have employed two techniques. One, in which we simulated the exact evolution of an ideal quantum computer. Since the cost of computing the exact evolution under a Hamiltonian scales exponentially with the number of qubits $n$, $\mathcal{O}(2^{3n})$, we have tested a less resource demanding method as well. By means of the first order Trotter expansion, we can approximate the evolution as
\begin{eqnarray}
    U_\text{T}&=&e^{-it\sum_{i<j}^n\sum_{\mu,\nu\in\{x,y,z\}}h_{i,j}^{\mu,\nu}\sigma^\mu_i\sigma^\nu_j}\nonumber\\
    &\approx& U_\text{appx}=\prod_{i<j}^n e^{-it\sum_{\mu,\nu\in\{x,y,z\}}h_{i,j}^{\mu,\nu}\sigma^\mu_i\sigma^\nu_j}\otimes\mathbb{1}_{r(i,j)},\nonumber\\
\end{eqnarray}
where $\mathbb{1}_{r(i,j)}$ is the identity matrix for the subspace of all qubits except $i$ and $j$. Using this approximation, we can reduce the cost of calculating the matrix exponential. Additionally, since every term is a sparse matrix with sparsity $\sim1-2^{-n}$, we can employ efficient functions for the matrix products. Techniques involving matrix-product-states (MPS) or matrix-product-operators (MPO) could be useful for extending this classical optimization strategy to larger systems~\cite{Vidal2003, GarciaMolina2023}.

\subsection{The parameter space}
The complete parameter space is given by the parameters of the SQG applied to qubit $i$ of the $k$-th analog block, $\{\theta_k^i, \phi_k^i$, $\lambda_k^i\}\in[0,2\pi)$, and the evolution time of the $k$-th block, $t_k$. We have restricted the evolution time of the analog blocks to be $0 \leq t_k \leq T\lVert H_\text{P}\rVert/K\lVert H_\text{S}\rVert$, where $K$ is the number of analog blocks, to avoid both negative and times much larger than the total time $T$ of the target evolution. As the number of qubits increases, this quickly leads to a wide and hard to explore parameter space, which makes convergence slow in the optimization process. We found a good compromise in reducing the kind of SQG applied on each block to just two, one $R_k(\theta_k, \phi_k$, $\lambda_k)$ applied to all even qubits and an additional $R'_k(\theta'_k, \phi'_k$, $\lambda'_k)$ applied to all odd qubits. These kind of odd-even SQG layers are the same as those required to compute the Trotterized evolution in many DAQC systems~\cite{STADAQC}, which gives a good starting point and sufficient flexibility for the optimization.

Calculating the evolution under the source Hamiltonian for all analog blocks is a resource-intensive task. As a mean of simplifying the computational cost and reducing the number of variables for the optimization, we have performed tests in which we fix the analog-block times to a fraction of the total evolution time of the target evolution $T$ divided by the total number of analog blocks $K$. Even though in certain cases one can achieve better results by having the evolution time $t_k$ as an optimization parameter, fixing $t_k = T/K$ results not only in a faster computation, but in a much faster and consistent convergence as well, as we show in Sec.~\ref{sec:example}.

\subsection{Optimization protocol}
Despite the compromises previously described, the optimization landscape is complex and shows multiple local minima. Moreover, evaluating the cost function for a new set of parameters is computationally very expensive. There are many strategies to tackle an optimization of this kind of ``black-box'' functions, such as genetic~\cite{GenAlg} or swarm~\cite{SwarmAlg} algorithms. 

In this work, we propose a combined strategy of Bayesian optimization and a gradient descent algorithm. The popular gradient descent consists of evaluating a point, computing the gradient of the function at that point and tuning the parameters following the slope of the function. It is a fast and efficient tool to exploit local minima, but falls short when searching for a global minimum and is highly dependent on the initial point of the optimization. Meanwhile, Bayesian optimization is specially advantageous to efficiently explore unknown and computationally expensive functions. Bayesian optimization treats the black-box function as a random function and considers a prior upon it. Then, it evaluates only the function, and not its derivative, to compute an acquisition function, which usually is based in the expected improvement or the probability of improvement. Finally, it uses this acquisition function to find the next point to evaluate and updates the prior for the next iteration. For an in depth read on Bayesian optimization, we refer the readers to Refs.~\cite{Bayesian_Mockus_1989,Bayesian_guide}.

As mentioned, the gradient descent is highly dependent on its initial point, as it usually converges towards the nearest local minimum, not the global one. A common way of dealing with this problem is performing several gradient descent optimization by changing the initial point, such as with random or grid searches on the parameter space~\cite{hypop}. The aforementioned Bayesian method allows us to minimize the number of initial points by means of a guided search through its acquisition function. The combined strategy leads to a much faster convergence to the global minimum than random or grid searches, achieving faster and better results.

\subsection{\label{sec:example}Example: Nearest-neighbour Hamiltonian}
A simple case of simulating the dynamics of the one-dimensional XY model is presented as a test for this strategy. Let us assume a target nearest neighbours Hamiltonian of the form
\begin{equation} \label{eq:test_target}
    H_\text{T} = g \sum_{i=1}^{n-1} \sigma^x_i \sigma^x_{i+1} + \sigma^y_i \sigma^y_{i+1}.
\end{equation}
Let's also assume that the source Hamiltonian to which we have access to is
\begin{equation} \label{eq:test_system}
    H_\text{S} = \sum_{i=1}^{n-1} h^{xz}_i \sigma^x_i \sigma^z_{i+1} + h^{zx}_i \sigma^z_i \sigma^x_{i+1} + h^{zz}_i \sigma^z_i \sigma^z_{i+1}.
\end{equation}
This problem was also studied within the DAQC paradigm in Ref.~\cite{Tasio2021DAQCcrossResonance}. 

In the following, we will discuss two cases: the homogeneous case, in which all couplings strengths are equal to the target Hamiltonian coupling strength, $h^k_i=h\ \forall i,k$, and the inhomogeneous case. For the inhomogeneous case, we propose a more realistic scenario in which the coupling strengths of the system follow a Gaussian distribution around the value of the target coupling strength $\tilde{h}^k_i=\mathcal{N}(g,\sigma)$, where  $\sigma$ is the variance of the distribution.

To simulate Eq.~\ref{eq:test_target} using Eq.~\ref{eq:test_system} in the homogeneous case, the general approach can be simplified into applying a single-qubit $\pi$ rotation around the $x$ axis to all the qubits, $R_x^{(n)}(\pi) =\otimes_{i=1}^n \sigma^x$, so that the crossed terms ($\sigma^x\sigma^z$ and $\sigma^z\sigma^x$) change signs. Then, since the rotated and system Hamiltonian do not commute, we employ the Trotter approximation
\begin{equation}\label{eq:test_rotation}
\left[\left(R_x^{(n)}(\pi)\right)^\dagger e^{-i \frac{H_S T}{2n_T}} R_x^{(n)}(\pi) e^{-i \frac{H_S T}{2n_{T}}}\right]^{n_T} \approx e^{-i H_{ZZ} T},
\end{equation}
where $H_{ZZ} = g\sum_{i=1}^{n-1} \sigma^z_i \sigma^z_{i+1}$ and $n_T$ is the number of Trotter steps. The second step consist in sandwiching the resulting $H_{ZZ}$ evolution in either a $R_x(\pi/2)$ or $R_y(\pi/2)$ to rotate all qubits from $\sigma^z_i$ to $\sigma^y_i$ or $\sigma^x_i$, respectively. This, of course, requires an additional Trotterization. This approximation can be made arbitrarily precise by increasing the number of Trotter steps at the cost of increasing the number of required digital-analog blocks (4 blocks per Trotter step).

As previously described, we can use two or more analog blocks sandwiched between arbitrary rotations on even and odd qubits, and find the optimal rotation angles to achieve the best approximation. Since the set of all arbitrary rotations includes the specific rotations used for the Trotter approximation, the optimization must result in values necessarily equal or better than the Trotter approximation for the same amount of blocks.

\begin{figure}[ht!]    
    \includegraphics[width=\linewidth]{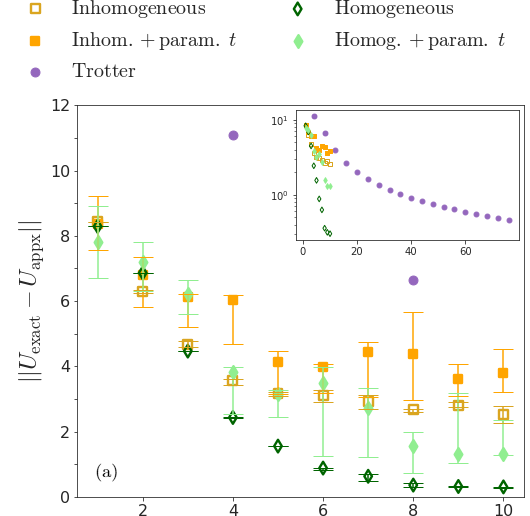}
    \includegraphics[width=\linewidth]{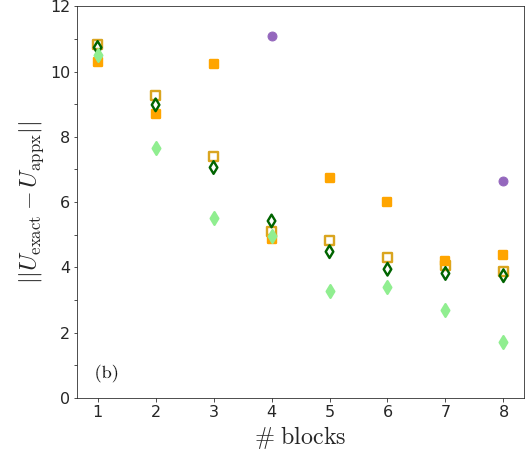}
    \caption{Frobenius distance between the optimized circuit and the exact evolution vs the number of digital-analog blocks. (a) We solve the problem of the nearest-neighbour Hamiltonian for 6 qubits. We show two cases, one in which the couplings in the system Hamiltonian is homogeneous, and the other where the coupling strength are in a normal distribution around the values of the former. We also distinguish two versions for the optimization process, which fixes the length of the analog blocks or leaves them as a variables. As a baseline, we represent the distances obtained with first order Trotterization up to $n_T = 19$ Trotter steps. The shown values indicate the mean Frobenius distance and the error bars show the interquartile range. The inset on top provides extended view in the coordinate axis for a higher number of blocks in logarithmic scale. (b) We repeat the procedure for a single run with a different classical simulation method, in which we employ a first order Trotter approximation for computing both the circuit and the target.}
    \label{fig:distance_comparison}
\end{figure}

In Fig.~\ref{fig:distance_comparison}, we show a comparison between the performance of a Trotter approximation and the proposed Bayesian plus gradient descent optimization with $N=6$ qubits. The chosen coupling strengths for the studied case where $h^k_i=g = 1\ \forall i,k$ in the homogeneous case and $\tilde{h}^k_i=\mathcal{N}(1,0.175)$ for the inhomogeneous one. Additionally, we performed both the optimization process with fixed evolution times and with $t_k$ as an optimization parameter were tested. For each point, we computed 20 runs, using only a 10-step optimization for the Gaussian process. Fixed-time optimizations achieve convergence in these 10 steps, as evidenced by their small error bars. Larger error bars show a higher variance in the achieved value for the cases with time as a parameter, meaning a lack of convergence and the need for a longer optimization with more steps to reach the global minimum. Comparing the results obtained with the Trotter formula we see an improvement on the Frobenius distance. In particular, we see an improvement of $\sim68\%$ and $\sim78\%$ for the 4 block cases in which the time is not a parameter for the inhomogeneous and homogeneous case respectively. When we include the analog block times as a parameter, we only obtain a mean improvement of $\sim45\%$ and $\sim65\%$ respectively. Although leaving the time as a parameter can lead to better results, the improved consistency and convergence rates of fixed-time optimization, along with a much lower computational cost, makes the fixed-time optimization the most desirable approach.

Homogeneous and random inhomogeneous couplings perform comparably. As expected, the case with homogeneous couplings have a slight advantage, as the target Hamiltonian is also homogeneous. This is good evidence that, as long as the coupling strengths associated with the system and the target Hamiltonians are similar, the approach is suitable for arbitrary two-body Hamiltonians.

Most importantly, all optimization approaches achieve a significantly smaller Frobenius distance, and therefore, a better fidelity than the traditional first order Trotter approximation for the same number of analog blocks. This way, by employing this classical optimization strategy on the digital layers (SQRs) of the simulation, one can achieve the same or better theoretical precision in the approximation saving up experimental resources. With current NISQ devices, this could directly lead to reduced experimental errors.

Additional non-extensive testing has been done in combining this optimization strategy with classical approximate simulations of the system. As the cost of classically calculating the unitary evolution under a Hamiltonian scales exponentially with the number of qubits, we need to find an alternative with a better scaling and which still can provide some advantage. With this goal, we employed the first order Trotter formula for simulating the analog blocks, which reduces the cost of the calculation when combined with sparse matrix multiplication algorithms. Here, instead of exactly calculating the evolution of our circuit, we employed the first order Trotter formula for approximating the evolution under both the target and the system Hamiltonians. These approximations, $U_\text{T,appx}$ and $U_\text{S,appx}$ respectively, are employed during the Bayesian optimization process. For checking the performance one would get when using the result of the optimization for running the quantum circuit, we calculate the Frobenius distance with the exact expressions. This way, we can directly compare the results obtained in both tests. We conclude that the Trotter approximation limits our capacity for optimizing the circuit. For the lowest amount of blocks for the Trotterization (4 blocks), we can obtain a $\sim54\%/51\%$ of improvement for the inhomogeneous/homogeneous problems when we discard the time as an optimizable parameter, and $\sim56\%/55\%$ improvement when we include it. This confirms the higher expresibility of the optimization model when we include the time as a parameter, at the cost of increasing the time to solution of the classical algorithm. However, for shallow circuits, the improvement of the fidelity is worth the optimization process. Particularly, the fixed-time optimization gives us a competitive Frobenius distance for circuits with 5 or less blocks compared to the Trotter decomposition. As it is also shown, we can simulate a system with a single analog block while maintaining the fidelity of the evolution obtained with 4 blocks employing the Trotter formula, $\lVert U_\text{exact}-U_\text{Trotter}\rVert=11.1$, $\lVert U_\text{exact}-U_\text{appx}\rVert=10.3$ for the inhomogeneous problem, and $\lVert U_\text{exact}-U_\text{appx}\rVert=10.5$ for the homogeneous.

\section{\label{sec:Conclusions}Conclusions}

In this article, we provide a set of versatile tools to implement DAQC circuits in realistic scenarios, where the terms in the Hamiltonians of both the system and the target do not commute by pairs. Following on the universality of DAQC, we have proposed a protocol for simulating an arbitrary two-body Hamiltonian by employing another arbitrary two-body Hamiltonian. This was achieved by sandwiching the analog blocks with pairs of SQGs picked from the Pauli gate set. This systematic method is universal, and provides an optimal number of digital-analog blocks $Q$ that scales quadratically with the number of qubits for all-to-all Hamiltonians, $Q\leq9n^2$. This strategy employs less blocks compared to previous approaches, which required $Q\leq36n^3$ steps. Additionally, we have proposed a new method for obtaining DAQC protocols in which all analog block times are positive, and thus, are implementable in practice.

With the same objective, we have explored a less resource-intensive approximation. We have proposed a classical strategy in which all possible SQG rotations are considered as digital blocks. Solving this optimization problem, we have shown that one can achieve an accuracy comparable or even better than large sequences of Trotter steps employing significantly less analog blocks. The benefits are twofold, as not only the accuracy of the systematic error is decresed, but a reduced amount of analog blocks also reduces the experimental error. In this regard, we have proposed a classical optimization strategy consisting in a Bayesian optimization method that explores the parameter space to find the optimal initial point of a gradient descent. This process is done without any assumption about the initial state, as the information given to the routine is just the target unitary evolution. For a low number of Trotter steps, we have shown that a precision comparable to that of regular Trotter approximations can be achieved by using only a fraction of the experimental resources.

The two approaches we have proposed in this work provide new tools for the implementation of DAQC circuits in NISQ devices. These new protocols for working with arbitrary two-body Hamiltonians pave the way for generating and scaling experimental implementations of this paradigm. The next steps for widening DAQC should search for an optimal solution for the negative analog block times problem, and further extend the protocols for $k$-body Hamiltonians. The exact method involve $\mathcal{O}(3^kn^2)$ digital-analog blocks in the worst case, but $\sim 9n^2$ on numerical simulations. This suggests that approximate solutions, similar to the one proposed in this article, should be explored for obtaining implementable DAQC circuits.

\begin{acknowledgments}
The authors thank M. Reichert, P. Rodr\'iguez and M. Capela for useful discussions and valuable feedback. ASC, PP-F and LL acknowledge financial support from the Spanish grant PID2019-104002GB-C21, PID2019-104002GB-C22 funded by Ministerio de Ciencia e Innovaci\'on/Agencia Estatal de Investigaci\'{o}n MCIN/AEI/10.13039/501100011033, FEDER ``A Way of Making Europe'', Consejer\'{\i}a de Conocimiento, Investigaci\'on y Universidad, Junta de Andaluc\'{\i}a, European Regional Development Fund (ERDF) under project US-1380840, grant Groups FQM-160, FQM-177, and the project PAIDI 2020 with reference P20\_01247 and P20\_00617 funded by the Consejer\'ia de Econom\'ia, Conocimiento, Empresas y Universidad, Junta de Andaluc\'ia (Spain). The authors aknowledge Resources supporting this work provided by the CEAFMC and Universidad de Huelva High Performance Computer (HPC@UHU) funded by ERDF/MI\-NE\-CO project UNHU-15CE-2848. MGdA and MS acknowledge support from EU FET Open project EPIQUS (899368) and HORIZON-CL4-2022-QUANTUM01-SGA project 101113946 OpenSuperQPlus100 of the EU Flagship on Quantum Technologies, the Spanish Ram\'on y Cajal Grant RYC-2020-030503-I, project Grant No. PID2021-125823NA-I00 funded by MCIN/AEI/10.13039/501100011033 and by “ERDF A way of making Europe” and “ERDF Invest in your Future”, and from the IKUR Strategy under the collaboration agreement between Ikerbasque Foundation and BCAM on behalf of the Department of Education of the Basque Government. MGdA acknowledges support from the UPV/EHU and TECNALIA 2021 PIF contract call. This work has been financially supported by the Ministry of Economic Affairs and Digital Transformation of the Spanish Government through the QUANTUM ENIA project call - Quantum Spain project, and by the European Union through the Recovery, Transformation and Resilience Plan - NextGenerationEU within the framework of the "Digital Spain 2026 Agenda". The code employed for the numerical experimentation is available under reasonable request.
\end{acknowledgments}

\appendix
\section{\label{Apx:universality}Universality of the protocol}

We want to solve the problem of simulating an arbitrary $n$ qubit two-body ATA Hamiltonian
\begin{equation}\label{eq:HT}
    H_\text{T}=\sum_{i<j}^n\sum_{\mu,\nu\in\{x,y,z\}}g_{i,j}^{\mu,\nu}\sigma^\mu_i\sigma^\nu_j,
\end{equation}
with a DAQC schedule employing another arbitrary two-body Hamiltonian,
\begin{equation}\label{eq:HS}
    H_\text{S}=\sum_{i<j}^n\sum_{\mu,\nu\in\{x,y,z\}}h_{i,j}^{\mu,\nu}\sigma^\mu_i\sigma^\nu_j,
\end{equation}
We are assuming that if we have a non-zero $g_{i,j}^{\mu,\nu}$ term, we will have a non zero $h_{i,j}^{\mu,\nu}$ coupling in the source Hamiltonian.

The protocol we propose is the following: select each of the possible pair of qubits $i,j$, and apply all the combinations of $\{x,y,z\}$ gates, which are the corresponding Pauli gates. This, changes the signs of the effective couplings, according to a $(\pm1)$ matrix, which we will call $M(n)$. Now, we want that our DAQC schedule simulates our original problem, this is
\begin{widetext}
\begin{equation}\label{eq:mainProblem}
    e^{-iTH_\text{T}}\approx\prod_{i<j}^{n}\prod_{\mu,\nu\in\{x,y,z\}}\exp\left(-it_{i,j}^{\mu,\nu}\sum_{i'<j'}^n\sum_{\mu',\nu'\in\{x,y,z\}}M(n)_{i,j,i',j'}^{\mu,\nu,\mu',\nu'}g_{i',j'}^{\mu',\nu'}\sigma^{\mu'}_{j'}\sigma^{\nu'}_{j'}\right).
\end{equation}
\end{widetext}
To prove that this protocol is universal we have to prove that we can always find a set of times for the duration of the analog blocks ($t_{i,j}^{\mu,\nu}$).

\subsection{Notation and definitions}
We define the notation employed for the proof. We employ a single index for labeling pairs of qubits,
\begin{equation}
    b_\ell(i,j,n)=n(i-1)-i(i+1)/2+j,\ 1\leq i<j\leq n.
\end{equation} 
Whenever we refer to a label of a coupling or a pair of qubits, we will be using this labeling convention. We also give a formula for the inverse of this indexing~\cite{APX:OEIS}
\begin{equation}
\begin{split}
    &i_\ell(b,n)=n-\left\lfloor\sqrt{n(n-1)-2b+2}+\frac{1}{2}\right\rfloor,\\
    &j_\ell(b,n)=b-n(i_\ell(b)-1)+\frac{1}{2}i_\ell(b)(i_\ell(b)+1).
\end{split}
\end{equation}
Additionally, we define an indexing method for the different selection of SQGs for a pair of qubits, $f_\ell(\mu,\nu)$. This indices are given from 1 to 9 according to the order in the following list: $\{xx,xy, xz,yx,yy,yz,zx,zy,zz\}$. For example, the pair of SQGs $\{y,z\}$ has the index $f_\ell(y,z)=6$. 

The elements $M(n)_{i,j,i',j'}^{\mu,\nu,\mu',\nu'}$ can be expressed as a matrix by employing the previous labeling of the pair of qubits, which can be then mapped to a matrix $M(n)_{i,j,i',j'}^{\mu,\nu,\mu',\nu'}=M(n)_{g_\ell(i',j',\mu',\nu'),g_\ell(i,j,\mu,\nu)}$. The rows and columns of this matrix are defined as
\begin{equation}\label{eq:rowcollabel}
    g_\ell(i,j,\mu,\nu)=9(b_\ell(i,j)-1)+f_\ell(\mu,\nu).
\end{equation}

The times for each analog block $t_{i,j}^{\mu,\nu}$ can be expressed as a column vector by employing the same labeling as in Eq.~\ref{eq:rowcollabel}, $t_{i,j}^{\mu,\nu}=t_{g_\ell(i,j,\mu,\nu)}$. We also define a new column vector $\overrightarrow{g/h}$, which contains the information about the couplings in the hardware and the couplings we want to simulate, again, we employ the same labeling as in Eq.~\ref{eq:rowcollabel} such that $(\overrightarrow{g/h})_{i,j}^{\mu,\nu}=(\overrightarrow{g/h})_{g_\ell(i,j,\mu,\nu)}$. Now, the problem of finding the times for the analog blocks can be expressed as 
\begin{equation}\label{eq:sysofeq}
    M(n)\overrightarrow{t}=T\overrightarrow{h/g}.
\end{equation}

Let's write the full $M(n)$ matrix as a block matrix,
\begin{equation}\label{eq:Mblocks}
    M(n)=\begin{pmatrix}
    A(n) & P(n)\\
    Q(n) & M(n-1)
    \end{pmatrix}
\end{equation}
where $A(n)$ is a $(n-1)\times(n-1)$ block matrix, and $M(n)$ a $n(n-1)/2\times n(n-1)/2$ block matrix. Here, each row labels a coupling in $H_\text{S}$ and each column labels the pair of qubits where the SQGs are applied. For example, the coupling between the qubits 1 and 4 would be addressed in row $I=b_\ell(1,4)=4$, and the blocks in which the SQGs are applied on qubits 2 and 3 correspond to the column $J=b_\ell(2,3)=n$. Each of these terms are defined also as block matrices (which we will call them from now on as sub-matrices). These blocks are $9\times9$ matrices, that can be one of the four $\{M^2,M^{1.1},M^{1.2},M^0\}$, according to the following formula
\begin{equation}
    M_{I,J}=\begin{cases}
    M^2 & \text{if } I=J,\\
    M^{1.1} & \text{if } i_\ell(I)=i_\ell(J) \text{ or } i_\ell(I)=j_\ell(J),\\
    M^{1.2} & \text{if } j_\ell(I)=i_\ell(J) \text{ or } j_\ell(I)=j_\ell(J),\\
    M^0 & \text{else}.
    \end{cases}
\end{equation}
For a detailed construction of a $M(n)$ matrix see Sec.~\ref{sec:explicitM}.

Let's define each of these sub-matrices:
\begin{itemize}
    \item $M^2$: this is the matrix that represent the signs of the effective couplings $\mu,\nu$ when we apply a SQG to each of the two qubits $i,j$. The columns represent each pair of gates, in the order given by $f_\ell(\mu,\nu)$. Equivalently, each column represents the sign of each effective coupling between the pair of qubits $i,j$, with the same order as the columns. Then, we have the following matrix,
    \begin{equation}
        M^2=\left(\begin{array}{ccc|ccc|ccc}
        1  & -1 & -1 & -1 & 1  & 1  & -1 & 1  & 1  \\
        -1 & 1  & -1 & 1  & -1 & 1  & 1  & -1 & 1  \\
        -1 & -1 & 1  & 1  & 1  & -1 & 1  & 1  & -1 \\ \hline
        -1 & 1  & 1  & 1  & -1 & -1 & -1 & 1  & 1  \\
        1  & -1 & 1  & -1 & 1  & -1 & 1  & -1 & 1  \\
        1  & 1  & -1 & -1 & -1 & 1  & 1  & 1  & -1 \\ \hline
        -1 & 1  & 1  & -1 & 1  & 1  & 1  & -1 & -1 \\
        1  & -1 & 1  & 1  & -1 & 1  & -1 & 1  & -1 \\
        1  & 1  & -1 & 1  & 1  & -1 & -1 & -1 & 1 
        \end{array}\right).
    \end{equation}
    \item $M^{1.1}$: in this case we only apply a SQG in the first qubit of the pair,
    \begin{equation}
        M^{1.1}=\left(\begin{array}{ccc|ccc|ccc}
        1  & 1  & 1  & -1 & -1 & -1 & -1 & -1 & -1 \\
        1  & 1  & 1  & -1 & -1 & -1 & -1 & -1 & -1 \\
        1  & 1  & 1  & -1 & -1 & -1 & -1 & -1 & -1 \\ \hline
        -1 & -1 & -1 & 1  & 1  & 1  & -1 & -1 & -1 \\
        -1 & -1 & -1 & 1  & 1  & 1  & -1 & -1 & -1 \\
        -1 & -1 & -1 & 1  & 1  & 1  & -1 & -1 & -1 \\ \hline
        -1 & -1 & -1 & -1 & -1 & -1 & 1  & 1  & 1  \\
        -1 & -1 & -1 & -1 & -1 & -1 & 1  & 1  & 1  \\
        -1 & -1 & -1 & -1 & -1 & -1 & 1  & 1  & 1
        \end{array}\right).
    \end{equation}
    \item $M^{1.2}$: the case we apply one SQG on the second qubit,
    \begin{equation}
        M^{1.2}=\left(\begin{array}{ccc|ccc|ccc}
        1  & -1 & -1 & 1  & -1 & -1 & 1  & -1 & -1 \\
        -1 & 1  & -1 & -1 & 1  & -1 & -1 & 1  & -1 \\
        -1 & -1 & 1  & -1 & -1 & 1  & -1 & -1 & 1  \\ \hline
        1  & -1 & -1 & 1  & -1 & -1 & 1  & -1 & -1 \\
        -1 & 1  & -1 & -1 & 1  & -1 & -1 & 1  & -1 \\
        -1 & -1 & 1  & -1 & -1 & 1  & -1 & -1 & 1  \\ \hline
        1  & -1 & -1 & 1  & -1 & -1 & 1  & -1 & -1 \\
        -1 & 1  & -1 & -1 & 1  & -1 & -1 & 1  & -1 \\
        -1 & -1 & 1  & -1 & -1 & 1  & -1 & -1 & 1 
         
        \end{array}\right).
    \end{equation}
    \item $M^0$: in this case neither SQGs is applied on any of the two qubits. This is the trivial case in which the signs of the effective couplings doesn't change,
    \begin{equation}
        M^0=\left(\begin{array}{ccc|ccc|ccc}
        1 & 1 & 1 & 1 & 1 & 1 & 1 & 1 & 1 \\
        1 & 1 & 1 & 1 & 1 & 1 & 1 & 1 & 1 \\
        1 & 1 & 1 & 1 & 1 & 1 & 1 & 1 & 1 \\ \hline
        1 & 1 & 1 & 1 & 1 & 1 & 1 & 1 & 1 \\
        1 & 1 & 1 & 1 & 1 & 1 & 1 & 1 & 1 \\ 
        1 & 1 & 1 & 1 & 1 & 1 & 1 & 1 & 1 \\ \hline
        1 & 1 & 1 & 1 & 1 & 1 & 1 & 1 & 1 \\
        1 & 1 & 1 & 1 & 1 & 1 & 1 & 1 & 1 \\
        1 & 1 & 1 & 1 & 1 & 1 & 1 & 1 & 1
        \end{array}\right).
    \end{equation}
\end{itemize}
For example, we apply an $x$ gate to the qubit 1 and an $y$ gate to the qubit 3. We want to know how does the sign of the coupling $yz$ changes between the qubits 2 and 3. In this case, we are applying a SQG only to the second qubit, so we have to look at the $M^{1.2}$ sub-matrix. Now, the SQG we have applied are $xy$, so this corresponds to the $2^\text{nd}$ column, and the $yz$ coupling to the $6$-th row. Looking at the matrix we see that the matrix, we see that in this case we have a change of sign of the effective coupling, and thus, we have a $-1$ in the corresponding matrix element.

We have some properties with these sub blocks: 
\begin{enumerate}[label=(S.\arabic*)]
    \item \label{S1} All the $\{M^2,M^{1.1},M^{1.2},M^0\}$ commute in pairs.
    \item \label{S2} $M^{1.1}M^{1.2}=M^2M^0=M^0$.
    \item \label{S3} $M^{1.1}M^0=M^{1.2}M^0=(-3)M^0$.
    \item \label{S4} $(M^0)^2=9M^0$.
    \item \label{S5} $(M^{1.1})^2=-3M^2M^{1.1}$.
    \item \label{S6} $(M^{1.2})^2=-3M^2M^{1.2}$.
\end{enumerate}

Also, we have some properties related to how these sub-blocks appear in the matrix:
\begin{enumerate}[label=(P.\arabic*)]
    \item \label{P1} $M(n)_{I,I} = M^2$.
    \item \label{P2} $M(n)_{I,J} = \{M^{1.1},M^{1.2},M^0\}, \forall I\neq J$.
    \item \label{P3} $M(n)_{I,J} = M^0 \Leftrightarrow M(n)_{J,I}=M^0$.
    \item \label{P4} $M(n)_{1,J} = \{M^2,M^{1.1},M^0\}$.
    \item \label{P5} $M(n)_{n(n-1)/2,J}= \{M^2,M^{1.2},M^0\}$.
    \item \label{P6} The number of sub-blocks that fulfill $M(n)_{I,J} = M(n)_{J,I} = M^{1.1}$ is the same as the number of sub-blocks fulfilling $M(n)_{I,J} = M(n)_{J,I} = M^{1.2}$.
\end{enumerate}

From now on, we will employ $N=n(n-1)/2$ to refer to the total number of pairs of qubits.

\subsection{Non-singularity of $M(n)$ through its determinant}
The main argument that we use for proving that our protocol is universal is that the matrix $M(n)$ that it generates is non-singular, and thus we can always obtain a set of times for the analog blocks that solves the equation in Eq.~\ref{eq:mainProblem}. We know that a matrix is non-singular if and only if its determinant is different from zero. Thus, we try to prove that $\text{Det}(M(n))=0$. For this, we employ the expansion of the formal determinant\footnote{The formal determinant has exactly the same expression as a usual determinant, but instead of the elements being numbers, they are square matrices of equal size that commute with each other by pairs.}~\cite{APX:Ingraham1937}, labeled with ``det'', of our block matrix,
\begin{equation}
    \text{Det}(M(n))=\text{Det}(\text{det}(M(n))).
\end{equation}

If we study the properties \ref{S1}-\ref{S6}, we see that if in a term of a determinant we have a $M^0$ element or a pair $M^{1.1}M^{1.2}$, that term will be proportional to $M^0$. The rest are cases in which we have terms with $\{M^2,M^{1.1}\}$, $\{M^2,M^{1.2}\}$, or $\{M^2\}$. By using the properties \ref{P1}-\ref{P6}, we see that the last case appears only once, and correspond to the elements of the main diagonal. Thus, we have that our formal determinant will be
\begin{eqnarray}\label{Eq:formalDet}
    \text{det}(M(n))&=&(M^2)^N+\alpha M^0\nonumber\\
    &\phantom{=}&+\sum_{k=2}^{N-1}\left[\beta_k(M^2)^{N-k}(M^{1.1})^k\right.\nonumber\\
    &\phantom{=}&\phantom{+}\left.+\gamma_k(M^2)^{N-k}(M^{1.2})^k\right],
\end{eqnarray}
where $\beta_k,\gamma_k\in\mathbb{Z}$. Looking at the properties \ref{S2}-\ref{S4}, we see that $\alpha=\pm3^q$, with $q\in\mathbb{N}$. 

Now, we will reorder the rows and columns. What we will do is to reorder them in such way that we obtain a expression equal to the one from Eq.~\ref{eq:Mblocks} but with the $M^{1.1}$ and $M^{1.2}$ swapped. In the original ordering, we wrote the $b$ indices appear in ascending order, such that the corresponding first label of the qubits appear in ascending order, and for the same first label, the second labels are in ascending order. Now, if we employ a new labels for the pairs which inverts all the ordering to be in descending order, $b'(i,j)=n(n-j)-(n-j+1)(n-j+2)/2+(n-i+1)$. Now, if we compute the formal determinant of the matrix with the new labels, we have
\begin{eqnarray}
    \text{det}(M'(n))&=&(M^2)^N+\alpha M^0\nonumber\\
    &\phantom{=}+&\sum_{k=2}^{N-1}\left[\beta'_k(M^2)^{N-k}(M^{1.2})^k\right.\nonumber\\
    &\phantom{=}&\phantom{+}\left.+\gamma'_k(M^2)^{N-k}(M^{1.1})^k\right]. 
\end{eqnarray}
Since we have performed as many changes in the rows as in the columns, the number of total swaps of rows and columns is even. Since the properties P1-4 are invariant under this change, we have that $\text{Det}(M(n))=\text{Det}(M'(n))\ \Rightarrow \text{det}(M(n))=\text{det}(M'(n))$. With this, we can conclude that $\beta_k=\beta'_k=\gamma_k=\gamma'_k$. For clarity, let's rewrite Eq.~\ref{Eq:formalDet} as
\begin{equation}
    \text{det}(M(n))=m(n)=(M^2)^N+\sum_{k=2}^{N-1}\beta_kB_k+\alpha M^0, 
\end{equation}
where we have evaluated the analytical expression for $B_k$ in \textit{Mathematica},
\begin{eqnarray}
    B_k&=&(M^2)^{N-k}(M^{1.1})^k+(M^2)^{N-k}(M^{1.2})^k\nonumber\\
    &=&\left(\begin{array}{ccc|ccc|ccc}
        \tilde{x}_k & \tilde{y}_k & \tilde{y}_k & \tilde{y}_k & \tilde{z}_k & \tilde{z}_k & \tilde{y}_k & \tilde{z}_k & \tilde{z}_k \\
        \tilde{y}_k & \tilde{x}_k & \tilde{y}_k & \tilde{z}_k & \tilde{y}_k & \tilde{z}_k & \tilde{z}_k & \tilde{y}_k & \tilde{z}_k \\
        \tilde{y}_k & \tilde{y}_k & \tilde{x}_k & \tilde{z}_k & \tilde{z}_k & \tilde{y}_k & \tilde{z}_k & \tilde{z}_k & \tilde{y}_k \\ \hline
        \tilde{y}_k & \tilde{z}_k & \tilde{z}_k & \tilde{x}_k & \tilde{y}_k & \tilde{y}_k & \tilde{y}_k & \tilde{z}_k & \tilde{z}_k \\
        \tilde{z}_k & \tilde{y}_k & \tilde{z}_k & \tilde{y}_k & \tilde{x}_k & \tilde{y}_k & \tilde{z}_k & \tilde{y}_k & \tilde{z}_k \\
        \tilde{z}_k & \tilde{z}_k & \tilde{y}_k & \tilde{y}_k & \tilde{y}_k & \tilde{x}_k & \tilde{z}_k & \tilde{z}_k & \tilde{y}_k \\ \hline
        \tilde{y}_k & \tilde{z}_k & \tilde{z}_k & \tilde{y}_k & \tilde{z}_k & \tilde{z}_k & \tilde{x}_k & \tilde{y}_k & \tilde{y}_k \\
        \tilde{z}_k & \tilde{y}_k & \tilde{z}_k & \tilde{z}_k & \tilde{y}_k & \tilde{z}_k & \tilde{y}_k & \tilde{x}_k & \tilde{y}_k \\
        \tilde{z}_k & \tilde{z}_k & \tilde{y}_k & \tilde{z}_k & \tilde{z}_k & \tilde{y}_k & \tilde{y}_k & \tilde{y}_k & \tilde{x}_k
        \end{array}\right)\nonumber\\
    &=&\left(\begin{array}{c|c|c}
        V_k & W_k & W_k \\ \hline
        W_k & V_k & W_k \\ \hline
        W_k & W_k & V_k \\ 
        \end{array}\right),
\end{eqnarray}
with the elements
\begin{equation}
\begin{split}
    &\tilde{x}_k=2(-1)^k3^{k-2}\left(1+(-1)^N2^{N+1}\right),\\
    &\tilde{y}_k=(-1)^k3^{k-2}\left(2+(-2)^N\right),\\
    &\tilde{z}_k=(-2)(-1)^k3^{k-2}\left(-1+(-2)^N\right).
\end{split}
\end{equation}
We also evaluate the expression for the powers of $M^2$,
\begin{equation}
    (M^2)^N=\left(\begin{array}{ccc|ccc|ccc}
        \tilde{a} & \tilde{b} & \tilde{b} & \tilde{b} & \tilde{c} & \tilde{c} & \tilde{b} & \tilde{c} & \tilde{c} \\
        \tilde{b} & \tilde{a} & \tilde{b} & \tilde{c} & \tilde{b} & \tilde{c} & \tilde{c} & \tilde{b} & \tilde{c} \\
        \tilde{b} & \tilde{b} & \tilde{a} & \tilde{c} & \tilde{c} & \tilde{b} & \tilde{c} & \tilde{c} & \tilde{b} \\ \hline
        \tilde{b} & \tilde{c} & \tilde{c} & \tilde{a} & \tilde{b} & \tilde{b} & \tilde{b} & \tilde{c} & \tilde{c} \\
        \tilde{c} & \tilde{b} & \tilde{c} & \tilde{b} & \tilde{a} & \tilde{b} & \tilde{c} & \tilde{b} & \tilde{c} \\
        \tilde{c} & \tilde{c} & \tilde{b} & \tilde{b} & \tilde{b} & \tilde{a} & \tilde{c} & \tilde{c} & \tilde{b} \\ \hline
        \tilde{b} & \tilde{c} & \tilde{c} & \tilde{b} & \tilde{c} & \tilde{c} & \tilde{a} & \tilde{b} & \tilde{b} \\
        \tilde{c} & \tilde{b} & \tilde{c} & \tilde{c} & \tilde{b} & \tilde{c} & \tilde{b} & \tilde{a} & \tilde{b} \\
        \tilde{c} & \tilde{c} & \tilde{b} & \tilde{c} & \tilde{c} & \tilde{b} & \tilde{b} & \tilde{b} & \tilde{a}
        \end{array}\right)=\left(\begin{array}{c|c|c}
        D & E & E \\ \hline
        E & D & E \\ \hline
        E & E & D \\ 
        \end{array}\right)
\end{equation}
with the elements
\begin{equation}
\begin{split}
    &\tilde{a}=3^{-2}\left(1+(-2)^{N+2}+2^{2N+2}\right),\\
    &\tilde{b}=3^{-2}\left(1+(-2)^N-2^{2N+1}\right),\\
    &\tilde{c}=3^{-2}\left(1+(-2)^{N+1}+2^{2N}\right).
\end{split}
\end{equation}
We can give a explicit expression for the matrix $m(n)$,
\begin{equation}
    m(n)=\left(\begin{array}{c|c|c}
        P & Q & Q \\ \hline
        Q & P & Q \\ \hline
        Q & Q & P \\ 
        \end{array}\right),
\end{equation}
where
\begin{equation}
\begin{split}
    &P=D+\sum_{k=2}^{N-1}\beta_kV_k+\alpha m^0,\\
    &Q=E+\sum_{k=2}^{N-1}\beta_kW_k+\alpha m^0,
\end{split}
\end{equation}
and
\begin{equation}
    m^0=\begin{pmatrix}
    1 & 1 & 1\\
    1 & 1 & 1\\
    1 & 1 & 1
    \end{pmatrix}.
\end{equation}

To evaluate the determinant of this matrix, since we have again a block structure. Checking that $[P,Q]=0$, we can employ the formal determinant trick to evaluate the determinant of the initial matrix. Evaluating this matrix, we obtain 
\begin{eqnarray}\label{eq:fullDet}
    \text{Det}(M(n))&=&\text{Det}(\text{det}(M(n)))=\text{Det}(m(n))\nonumber\\
    &=&\text{Det}(\text{det}(m(n))=\text{Det}(P^3-3PQ^2+2Q^3)\nonumber\\
    &=&(p+q-2r)^4(p-2q+r)^4(p+4q+4r),\nonumber\\
\end{eqnarray}
with the elements
\begin{eqnarray}
    p&=&3^{-2}\left[1+(-2)^{N+2}+2^{2N+2}+(-3)^{q+2}\right.\nonumber\\
    &\phantom{=}&\left.+\sum_{k=2}^{N-1}\beta_k2(-3)^k\left(1+(-1)^N2^{N+1}\right)\right],\nonumber\\
    q&=&3^{-2}\left[1+(-2)^N-2^{2N+1}+(-3)^{q+2}\right.\nonumber\\
    &\phantom{=}&\left.+\sum_{k=2}^{N-1}\beta_k(-3)^k\left(2+(-2)^N\right)\right],\nonumber\\
    r&=&3^{-2}\left[1+(-2)^{N+2}+2^{2N+2}+(-3)^{q+2}\right.\nonumber\\
    &\phantom{=}&\left.+\sum_{k=2}^{N-1}\beta_k(-2)(-3)^k\left(-1+(-2)^N\right)\right],
\end{eqnarray}
where we have already substituted the expression for $\alpha=(-3)^q$, $q\in\mathbb{N}$.

\subsection{Modular arithmetic for proving that $\text{Det}(M(n))\neq0$}
To prove that the determinant is not zero, we have to prove that each of the expressions for the determinant in Eq.~\ref{eq:fullDet} are not zero. For this, we will employ a result from modular arithmetic for each of the terms:

\begin{enumerate}[label=(T.\arabic*)]
\item This term can be written as
\begin{eqnarray}
    (p+q-2r)&=&3^{-2}\left[2^{2N+2}+3^2(-2)^N+\right.\nonumber\\
    &\phantom{=}&\left.+\sum_{k=2}^{N-1}\beta_k(-2)^N(-3)^{k+2}\right].
\end{eqnarray}
A fundamental property from modular algebra is that for an integer $a$ to be equal to 0, we have the necessary condition $a\mod b=0$, $\forall b\in\mathbb{N}$. In particular, we can check this condition for the modulo 3,
\begin{eqnarray}
    &&3^2(p+q-2r)\text{mod }3\nonumber\\
    &&= \left[2^{2N+2}+3^2(-2)^N+\sum_{k=2}^{N-1}\beta_k(-2)^N(-3)^{k+2}\right]\text{mod }3\nonumber\\
    &&= \left[ 2^{2N+2} \text{mod } 3 + 3^2(-2)^N \text{mod } 3 \right.\nonumber\\
    &&\phantom{=}\left. + \left(\sum_{k=2}^{N-1}\beta_k(-2)^N(-3)^{k+2}\right)\text{mod }3\right]\text{mod }3\nonumber\\
    &&= \left[ 2^{2N+2} \text{mod } 3 + 0 + 0 \right]\text{mod }3\nonumber\\
    &&=1 \neq 0,
\end{eqnarray}
where we have employed the compatibility with addition\footnote{$(a+b)\mod x=(a\mod x+b\mod x)\mod x$} and multiplication\footnote{$ab\mod x=[(a\mod x)(b\mod x)]\mod x$}, and the property that $a^k\mod a=0$, $\forall k\in\mathbb{N}$.

\item The second term is trivially not zero, since
\begin{equation}
    (p-2q+r)=3^{-2}4^N\neq0.
\end{equation}

\item For the third term we employ a similar proof as for the first one,
\begin{eqnarray}
    &&3^2(p+4q+4r)\text{mod }3\nonumber\\
    &&= \left[3^2+(-3)^{4+q}+2^{2N+4}+\sum_{k=2}^{N-1}\beta_k2(-3)^{k+2}\right]\text{mod }3\nonumber\\
    &&=\left[3^2\text{mod }3+(-3)^{4+q}\text{mod }3+2^{2N+4}\text{mod }3\right.\nonumber\\
    &&\phantom{=}\left.+\left(\sum_{k=2}^{N-1}\beta_k2(-3)^{k+2}\right)\text{mod }3\right]\text{mod }3\nonumber\\
    &&=\left[0+0+1+0\right]\text{mod }3\nonumber\\
    &&=1\neq0.
\end{eqnarray}
\end{enumerate}

Thus, we have proved that the determinant of $M(n)$ is not zero, so that it is invertible, and thus, we can find a solution for solving the times for the analog blocks in this schedule. $\blacksquare$

\subsection{\label{sec:explicitM}Detailed construction of the $M(n)$ matrix}

As we have seen in the previous section, we can build the matrix $M(n)$ by directly using the definition of the $\{M^0,M^{1.1},M^{1.2},M^2\}$ blocks. Also, we can build it iteratively from all the $M(n)$ matrices of smaller sizes. For simplicity, we will employ the second method. 

Let us start by assuming that we have already the $M(n-1)$ matrix. Now, we need to construct a explicit way for the $A(n)$, $P(n)$, and $Q(n)$ matrices from Eq.~\ref{eq:Mblocks}.

For the $(n-1)\times(n-1)$ block matrix $A(n)$ we have an easy structure, with all the sub-blocks outside the main diagonal being $M^{1.1}$ and the elements in the diagonal being $M^2$, 
\begin{equation} 
    A(n)=\begin{pmatrix}
    M^2 & M^{1.1} & \dots & M^{1.1} \\
    M^{1.1} & M^2 & \ddots & \vdots \\
    \vdots & \ddots & \ddots & M^{1.1}\\
    M^{1.1} & \dots & M^{1.1} & M^2
    \end{pmatrix}. 
\end{equation}

For the $P(n)$ matrix we know that the only sub-matrices that can appear are $M^0$ and $M^{1.2}$. This property comes from the fact that all the couplings that the matrix $P(n)$ addresses have the first qubit in common, thus the pairs of qubits can only be equal to the second qubit or none. This makes easier to identify the blocks in this matrix. Employing the $I,J$ indices from Eq.~\ref{eq:Mblocks} we have that
\begin{equation}
    P(n)=M_{I^*,J^*}=\begin{cases}
    M^{1.2} & \text{if } j(I^*)=j(J^*) \vee j(I^*)=i(J^*),\\
    M^0 &\text{else},
    \end{cases}
\end{equation}
with the indices restricted to $1\leq I^*\leq n-1$ and $n\leq J^* \leq n(n-1)/2$.

The $Q(n)$ matrix is defined in a similar manner, but here we also have $M^{1.1}$ blocks,
\begin{equation}
    Q(n)=M_{I^*,J^*}=\begin{cases}
    M^{1.1} & \text{if } i(I^*)=j(J^*),\\
    M^{1.2} & \text{if } j(I^*)=j(J^*),\\
    M^0 &\text{else},
    \end{cases}
\end{equation}
with the indices restricted to $n\leq I^*\leq n(n-1)/2$ and $1\leq J^* \leq n-1$.\newline

With these definitions one can construct the $M(n)$ matrix by starting from the smallest possible matrix $M(2)$ and building the matrix with one more qubit. As an example, we build the matrix for up to $n=5$ qubits,

\begin{widetext}
\begin{eqnarray}
    M(2) = \left( M^2\right),\nonumber\\
    M(3) = \left(\begin{array}{lll}
        M^2 & M^{1.1} & M^{1.2}\\
        M^{1.1} & M^2 & M^{1.2}\\\cline{3-3} 
        M^{1.1} & \multicolumn{1}{l|}{M^{1.2}} & M^2
        \end{array}\right),\nonumber\\
    M(4) = \left(\begin{array}{llllll}
        M^2     & M^{1.1} & M^{1.1}                      & M^{1.2} & M^{1.2}                      & M^0     \\
        M^{1.1} & M^2     & M^{1.1}                      & M^{1.2} & M^0                          & M^{1.2} \\
        M^{1.1} & M^{1.1} & M^2                          & M^0     & M^{1.2}                      & M^{1.2} \\ \cline{4-6} 
        M^{1.1} & M^{1.2} & \multicolumn{1}{l|}{M^0}     & M^2     & M^{1.1}                      & M^{1.2} \\
        M^{1.1} & M^0     & \multicolumn{1}{l|}{M^{1.2}} & M^{1.1} & M^2                          & M^{1.2} \\ \cline{6-6} 
        M^0     & M^{1.1} & \multicolumn{1}{l|}{M^{1.2}} & M^{1.1} & \multicolumn{1}{l|}{M^{1.2}} & M^2    
        \end{array}\right),\nonumber\\
    M(5) = \left(\begin{array}{llllllllll}
        M^2     & M^{1.1} & M^{1.1} & M^{1.1}                      & M^{1.2} & M^{1.2} & M^{1.2}                      & M^0     & M^0                          & M^0     \\
        M^{1.1} & M^2     & M^{1.1} & M^{1.1}                      & M^{1.2} & M^0     & M^0                          & M^{1.2} & M^{1.2}                      & M^0     \\
        M^{1.1} & M^{1.1} & M^2     & M^{1.1}                      & M^0     & M^{1.2} & M^0                          & M^{1.2} & M^0                          & M^{1.2} \\
        M^{1.1} & M^{1.1} & M^{1.1} & M^2                          & M^0     & M^0     & M^{1.2}                      & M^0     & M^{1.2}                      & M^{1.2} \\ \cline{5-10} 
        M^{1.1} & M^{1.2} & M^0     & \multicolumn{1}{l|}{M^0}     & M^2     & M^{1.1} & M^{1.1}                      & M^{1.2} & M^{1.2}                      & M^0     \\
        M^{1.1} & M^0     & M^{1.2} & \multicolumn{1}{l|}{M^0}     & M^{1.1} & M^2     & M^{1.1}                      & M^{1.2} & M^0                          & M^{1.2} \\
        M^{1.1} & M^0     & M^0     & \multicolumn{1}{l|}{M^{1.2}} & M^{1.1} & M^{1.1} & M^2                          & M^0     & M^{1.2}                      & M^{1.2} \\ \cline{8-10} 
        M^0     & M^{1.1} & M^{1.2} & \multicolumn{1}{l|}{M^0}     & M^{1.1} & M^{1.2} & \multicolumn{1}{l|}{M^0}     & M^2     & M^{1.1}                      & M^{1.2} \\
        M^0     & M^{1.1} & M^0     & \multicolumn{1}{l|}{M^{1.2}} & M^{1.1} & M^0     & \multicolumn{1}{l|}{M^{1.2}} & M^{1.1} & M^2                          & M^{1.2} \\ \cline{10-10} 
        M^0     & M^0     & M^{1.1} & \multicolumn{1}{l|}{M^{1.2}} & M^0     & M^{1.1} & \multicolumn{1}{l|}{M^{1.2}} & M^{1.1} & \multicolumn{1}{l|}{M^{1.2}} & M^2    
        \end{array}\right).
\end{eqnarray}
\end{widetext}

\section{\label{Apx:PositiveTimes}Obtaining positive times for the DA schedule}

In this appendix, we will prove that we we can always find a DAQC schedule for simulating an arbitrary ATA two-body Hamiltonian with another one in which the analog block times are positive. For this, we will first modify the problem such that we rewrite it as a non-negative least squares (NNLS) problem. Then, we will employ the Algorithm NNLS for solving the modified problem, which has a proven convergence to the optimal solution~\cite{APX:Lawson1974}. However, for proving that Algorithm NNLS will converge to a correct solution, we have to then prove that there exists at least one positive solution. 

\subsection{Algorithm NNLS}

The problem we want to solve is similar to the one in Eq.~\ref{eq:sysofeq} but for an arbitrary protocol $M$
\begin{equation}\label{eq:problemequation}
        M\vec{t}=T\overrightarrow{g/h}\equiv\vec{b},\text{ subject to }\vec{t}\geq0,
\end{equation}
where the column vector $\vec{b}\in\mathbb{R}^{(9n(n-1)/2)}$ with $n$ the number of qubits. From now on, we will employ $N=9n(n-1)/2$ to denote the dimension of the problem. The vector $b$ has all the information about both the target and the source Hamiltonians, and the simulation time. Within this subsection, we will assume that the matrix $M$ allows for a positive solution for any $\vec{b}$, $\vec{t}\geq0\ \forall i$. 

The NNLS problem consist in the minimization task $\lVert E\vec{x}-\vec{f}\rVert$, with $E\in\mathcal{M}_{N,m}(\mathbb{R})$ and $\vec{f}\in R^N$, subject to $x_i\geq0$. We can transform our problem in Eq.~\ref{eq:problemequation} to a NNLS by just substituting $E\rightarrow M$, $\vec{x}\rightarrow\vec{t}$ and $\vec{f}\rightarrow\vec{b}$. For solving our original problem we need to find a solution such that $\lVert M\vec{t}-\vec{b}\rVert=0$.

For solving this problem, we will employ the Algorithm NNLS (see~\cite{APX:Lawson1974} for the details of the algorithm). A key characteristic of this algorithm is that it is proven that it converges to the optimal solution in a finite number of steps. Thus, it is direct to see that if there exist a solution to our original problem, there would also be a solution for its NNLS version with $\lVert M\vec{t}-\vec{b}\rVert=0$. 

We have run several numerical experiments for uniformly random problem vectors $\vec{b}$. We see that when employing this algorithm, the number of nonzero times in the output coincides with the number of coupling terms in the system.

\subsection{Constructive method for obtaining a positive solution for every problem}

The proof that the algorithm NNLS converges to a positive solution relies on the fact that such solution exists. In this subsection we will construct a general method for building positive solutions for this problem for all cases. With this we will prove that we can construct a DAQC protocol which flips the effective signs of the Hamiltonian terms such that the times of the analog blocks are always positive. 

As a sketch of the proof, we will start by noting that all possible DAQC schedules generates up to $4^n$ different columns to choose from to build the matrix $M$. Then, we note that these columns corresponds to the vertices of a polytope with center of mass in the origin of coordinates. We will prove that we can find a sum of the coordinates of these vertices with positive coefficients. Finally, we will show that projecting the coordinates of a small hyper-sphere around the center of mass we can write any other point in the hyperspace with a positive sum, and thus we can solve any problem in the same manner.

\subsubsection{Proof of the existence of a schedule with positive times}
Starting from the the general case of arbitrary two-body Hamiltonians, we need to find a protocol for simulating a a Hamiltonian consisting of $N$ two-body terms employing the same number of terms. As we did for the new protocol, we can restrict ourselves to sandwiching each analog block only with Pauli gates. These gates plus the identity, $\{\mathbb{1},X,Y,Z\}$, generate a pool of $4^n$ different combinations. In turn, these combinations of gates generates the same number of possible columns for the matrix $M$, $M=\{M_i\}_{i=1}^{4^n}$ such that $M_i\in\{\pm1\}^N$.

The rules to define a column $M_i$ given a selection of SQGs can be addressed qubit by qubit and coupling by coupling, employing the following relations $YXY=ZXZ=-X$, $XYX=ZYZ=-Y$, $XZX=YZY=-Z$ and $\mu\mu\mu=\mu$ for $\mu=\{\mathbb{1},X,Y,Z\}$. From these properties, one can note that each of the couplings $\sigma_i^\mu\sigma_j^\nu$ can only change if we apply a gate, i.e. $\{X,Y,Z\}$, to at least one of the two qubits $i$ or $j$.

Each selection of SQGs for the sandwiching will generate a new column $M_i$ which is linearly independent to every other column from the set $M$, $M_i\nparallel M_j\ \forall i<j\leq4^n$. As we have proven in Appendix~\ref{Apx:universality}, there is at least one combination of columns which forms a basis for a $N$-dimensional space. This assures that taking any other extra column, we can define a $N$-polytope. We define the convex $N$-polytope $\mathcal{P}$ as the volume with vertices in $M_i$, $\mathcal{P}=\{\vec{p}\in\mathbb{R}^N: \vec{p}=\sum_{i=1}^{4^n}\alpha_iM_i,\ \sum_{i=1}^{4^n}\alpha_i\leq1, \alpha_i\geq0\ \forall i\}$. 

Now we will employ a generalization of the theorem from Ref.~\cite{APX:Dobbins2015}, which states the following:
\begin{theorem}[\cite{APX:Blagojevic2015}]\label{th:polytope}
    Let $P$ be a $d$-plytope, $p\in P$, and $k$ and $n$ positive integers with $kn\geq d$. Then, there are points $p_1,\dots,p_n$ in the $k$-skeleton of $P$ with barycenter $p=\frac{1}{n}(p_1+\dots+p_n)$. 
\end{theorem}
For our proof, we will particularize this theorem to the $0$-skeleton by taking the extreme points of $4^n$ edges of the $1$-skeleton $p_1,\dots,p_{4^n}$, such that these points exactly coincide with the 0-skeleton, $\{p_1,\dots,p_{4^n}\}=\text{skel}_0(\mathcal{P})$. Now, we will prove that the barycenter of the 0-skeleton corresponds to the point $\vec{0}=(0,\dots,0)$, this is, the next step is to prove that
\begin{equation}\label{eq:sumapx}
    \sum_{i=1}^{4^n}M_i=\begin{pmatrix}
        0\\
        \vdots\\
        0
    \end{pmatrix}.
\end{equation}

We will start by focusing on a single row, the one corresponding of the $Z_1Z_2$ coupling. As stated previously, this coupling will change in a way that only takes into account the gates applied to qubits 1 and 2. The combination of gates that flips the sign of this coupling is $\{\mathbb{1}_1X_2,\mathbb{1}_1Y_2,Z_1X_2,Z_1Y_2,X_1\mathbb{1}_2,Y_1\mathbb{1}_2,X_1Z_2,Y_1Z_2\}$. The SQGs for the rest of the qubits can be chosen arbitrarily, so we have a total of $8\cdot4^{n-2}$ different columns in which the sign flips. As this number is exactly half of the total number of columns, the sum in Eq.~\ref{eq:sumapx} for the row corresponding to the $Z_1Z_2$ coupling is 0. This can be straightforwardly extended to all couplings repeating the same argument. Employing Th.~\ref{th:polytope} and Eq.~\ref{eq:sumapx}, we find that the barycenter of $\text{skel}_0(\mathcal{P})$ is the point $\vec{0}=4^{-n}\sum_{i=1}^{4^n}M_i$.

Now, we will construct a hypersphere $\mathcal{S}$ with small radius $r\ll4^{-n}$ centered in the point $\vec{0}$. For this, we will slightly distort the sum with an extra term $\vec{\varepsilon}=(\varepsilon_1,\dots,\varepsilon_{4^n})$ such that this vector is in the surface of $\mathcal{S}$, $\vec{s}=4^{-n}\vec{1}+\vec{\varepsilon}\in\mathcal{S}$ with $\sum_{i=1}^{4^n}\varepsilon_i=0$, $\lvert\sum_{i=1}^{4^n}\varepsilon_iM_i\rvert=r$ and $\lvert\varepsilon_i\rvert\leq4^{-n}\ \forall i$. Since we are adding a small deviation around the barycenter, the points in the surface of $\mathcal{S}$ are inside the polytope, $\mathcal{S}\in\mathcal{P}$. 

By definition of the hypersphere centered at the origin, any point $\vec{x}$ can be obtained by projecting a point of the hypersphere $\vec{s}$, $\exists!\vec{s}\in\mathcal{S}:\vec{x}=\lambda\vec{s},\ \lambda\geq0,\ \forall\vec{x}\in\mathbb{R}^N$. We can then write the original problem using this projection from the hypersphere to solve the problem in Eq.~\ref{eq:problemequation} with positive solutions,
\begin{eqnarray}
    \vec{b}&=&\lambda\vec{s}=\lambda\left[\sum_{i=1}^{4^n}(4^{-n}+\varepsilon_i)M_i\right]\nonumber\\
    &=& M\lambda\begin{pmatrix}
        4^{-n}+\varepsilon_1\\
        \vdots\\
        4^{-n}+\varepsilon_{4^n}
    \end{pmatrix}=M\vec{t}.
\end{eqnarray}
By construction, all the times of the analog blocks are positive, $t_i=\lambda(4^{-n}+\varepsilon_i)\geq0\ \forall i$. with this, we have proven that there exist at least one positive solution to the original problem, so that we can always converge to a positive solution by employing the Algorithm NNLS. $\blacksquare$

\bibliography{main.bib}

\begin{thebibliography}{38}%
\makeatletter
\providecommand \@ifxundefined [1]{%
 \@ifx{#1\undefined}
}%
\providecommand \@ifnum [1]{%
 \ifnum #1\expandafter \@firstoftwo
 \else \expandafter \@secondoftwo
 \fi
}%
\providecommand \@ifx [1]{%
 \ifx #1\expandafter \@firstoftwo
 \else \expandafter \@secondoftwo
 \fi
}%
\providecommand \natexlab [1]{#1}%
\providecommand \enquote  [1]{``#1''}%
\providecommand \bibnamefont  [1]{#1}%
\providecommand \bibfnamefont [1]{#1}%
\providecommand \citenamefont [1]{#1}%
\providecommand \href@noop [0]{\@secondoftwo}%
\providecommand \href [0]{\begingroup \@sanitize@url \@href}%
\providecommand \@href[1]{\@@startlink{#1}\@@href}%
\providecommand \@@href[1]{\endgroup#1\@@endlink}%
\providecommand \@sanitize@url [0]{\catcode `\\12\catcode `\$12\catcode
  `\&12\catcode `\#12\catcode `\^12\catcode `\_12\catcode `\%12\relax}%
\providecommand \@@startlink[1]{}%
\providecommand \@@endlink[0]{}%
\providecommand \url  [0]{\begingroup\@sanitize@url \@url }%
\providecommand \@url [1]{\endgroup\@href {#1}{\urlprefix }}%
\providecommand \urlprefix  [0]{URL }%
\providecommand \Eprint [0]{\href }%
\providecommand \doibase [0]{https://doi.org/}%
\providecommand \selectlanguage [0]{\@gobble}%
\providecommand \bibinfo  [0]{\@secondoftwo}%
\providecommand \bibfield  [0]{\@secondoftwo}%
\providecommand \translation [1]{[#1]}%
\providecommand \BibitemOpen [0]{}%
\providecommand \bibitemStop [0]{}%
\providecommand \bibitemNoStop [0]{.\EOS\space}%
\providecommand \EOS [0]{\spacefactor3000\relax}%
\providecommand \BibitemShut  [1]{\csname bibitem#1\endcsname}%
\let\auto@bib@innerbib\@empty
\bibitem [{\citenamefont {Benioff}(1980)}]{Benioff1980}%
  \BibitemOpen
  \bibfield  {author} {\bibinfo {author} {\bibfnamefont {P.}~\bibnamefont
  {Benioff}},\ }\href {https://doi.org/10.1007/BF01011339} {\bibfield
  {journal} {\bibinfo  {journal} {J. Stat. Phys.}\ }\textbf {\bibinfo {volume}
  {22}},\ \bibinfo {pages} {563} (\bibinfo {year} {1980})}\BibitemShut
  {NoStop}%
\bibitem [{\citenamefont {Feynman}(1982)}]{Feynman}%
  \BibitemOpen
  \bibfield  {author} {\bibinfo {author} {\bibfnamefont {R.~P.}\ \bibnamefont
  {Feynman}},\ }\href {https://doi.org/10.1007/BF02650179} {\bibfield
  {journal} {\bibinfo  {journal} {Int. J. Theor. Phys.}\ }\textbf {\bibinfo
  {volume} {21}},\ \bibinfo {pages} {467} (\bibinfo {year} {1982})}\BibitemShut
  {NoStop}%
\bibitem [{\citenamefont {Das}\ and\ \citenamefont
  {Chakrabarti}(2008)}]{Arnab2008}%
  \BibitemOpen
  \bibfield  {author} {\bibinfo {author} {\bibfnamefont {A.}~\bibnamefont
  {Das}}\ and\ \bibinfo {author} {\bibfnamefont {B.~K.}\ \bibnamefont
  {Chakrabarti}},\ }\href {https://doi.org/10.1103/RevModPhys.80.1061}
  {\bibfield  {journal} {\bibinfo  {journal} {Rev. Mod. Phys.}\ }\textbf
  {\bibinfo {volume} {80}},\ \bibinfo {pages} {1061} (\bibinfo {year}
  {2008})}\BibitemShut {NoStop}%
\bibitem [{\citenamefont {Deutsch}\ \emph {et~al.}(1995)\citenamefont
  {Deutsch}, \citenamefont {Barenco},\ and\ \citenamefont
  {Ekert}}]{Deutsch1995}%
  \BibitemOpen
  \bibfield  {author} {\bibinfo {author} {\bibfnamefont {D.~E.}\ \bibnamefont
  {Deutsch}}, \bibinfo {author} {\bibfnamefont {A.}~\bibnamefont {Barenco}},\
  and\ \bibinfo {author} {\bibfnamefont {A.}~\bibnamefont {Ekert}},\ }\href
  {https://doi.org/10.1098/rspa.1995.0065} {\bibfield  {journal} {\bibinfo
  {journal} {Proc. Roy. Soc. Lond. A Math.}\ }\textbf {\bibinfo {volume}
  {449}},\ \bibinfo {pages} {669} (\bibinfo {year} {1995})}\BibitemShut
  {NoStop}%
\bibitem [{\citenamefont {Werschnik}\ and\ \citenamefont
  {Gross}(2007)}]{Werschnik2007}%
  \BibitemOpen
  \bibfield  {author} {\bibinfo {author} {\bibfnamefont {J.}~\bibnamefont
  {Werschnik}}\ and\ \bibinfo {author} {\bibfnamefont {E.~K.~U.}\ \bibnamefont
  {Gross}},\ }\href {https://arxiv.org/abs/0707.1883} {\  (\bibinfo {year}
  {2007})},\ \Eprint {https://arxiv.org/abs/arXiv:0707.1883}
  {arXiv:arXiv:0707.1883 [quant-ph]} \BibitemShut {NoStop}%
\bibitem [{\citenamefont {Koch}\ \emph {et~al.}(2022)\citenamefont {Koch},
  \citenamefont {Boscain}, \citenamefont {Calarco}, \citenamefont {Dirr},
  \citenamefont {Filipp}, \citenamefont {Glaser}, \citenamefont {Kosloff},
  \citenamefont {Montangero}, \citenamefont {Schulte-Herbr{\"u}ggen},
  \citenamefont {Sugny},\ and\ \citenamefont {Wilhelm}}]{Koch2022}%
  \BibitemOpen
  \bibfield  {author} {\bibinfo {author} {\bibfnamefont {C.~P.}\ \bibnamefont
  {Koch}}, \bibinfo {author} {\bibfnamefont {U.}~\bibnamefont {Boscain}},
  \bibinfo {author} {\bibfnamefont {T.}~\bibnamefont {Calarco}}, \bibinfo
  {author} {\bibfnamefont {G.}~\bibnamefont {Dirr}}, \bibinfo {author}
  {\bibfnamefont {S.}~\bibnamefont {Filipp}}, \bibinfo {author} {\bibfnamefont
  {S.~J.}\ \bibnamefont {Glaser}}, \bibinfo {author} {\bibfnamefont
  {R.}~\bibnamefont {Kosloff}}, \bibinfo {author} {\bibfnamefont
  {S.}~\bibnamefont {Montangero}}, \bibinfo {author} {\bibfnamefont
  {T.}~\bibnamefont {Schulte-Herbr{\"u}ggen}}, \bibinfo {author} {\bibfnamefont
  {D.}~\bibnamefont {Sugny}},\ and\ \bibinfo {author} {\bibfnamefont {F.~K.}\
  \bibnamefont {Wilhelm}},\ }\href
  {https://doi.org/10.1140/epjqt/s40507-022-00138-x} {\bibfield  {journal}
  {\bibinfo  {journal} {EPJ Quantum Technol.}\ }\textbf {\bibinfo {volume}
  {9}},\ \bibinfo {pages} {19} (\bibinfo {year} {2022})}\BibitemShut {NoStop}%
\bibitem [{\citenamefont {Nielsen}\ and\ \citenamefont
  {Chuang}(2010)}]{NielsenChuang2010}%
  \BibitemOpen
  \bibfield  {author} {\bibinfo {author} {\bibfnamefont {M.~A.}\ \bibnamefont
  {Nielsen}}\ and\ \bibinfo {author} {\bibfnamefont {I.~L.}\ \bibnamefont
  {Chuang}},\ }\href {https://doi.org/10.1119/1.1463744} {\emph {\bibinfo
  {title} {Quantum Computation and Quantum Information}}}\ (\bibinfo
  {publisher} {Cambridge University Press},\ \bibinfo {year}
  {2010})\BibitemShut {NoStop}%
\bibitem [{\citenamefont {Kitaev}(1997)}]{Kitaev1997}%
  \BibitemOpen
  \bibfield  {author} {\bibinfo {author} {\bibfnamefont {A.~Y.}\ \bibnamefont
  {Kitaev}},\ }\bibinfo {title} {Quantum error correction with imperfect
  gates},\ in\ \href {https://doi.org/10.1007/978-1-4615-5923-8_19} {\emph
  {\bibinfo {booktitle} {Quantum Communication, Computing, and Measurement}}},\
  \bibinfo {editor} {edited by\ \bibinfo {editor} {\bibfnamefont
  {O.}~\bibnamefont {Hirota}}, \bibinfo {editor} {\bibfnamefont {A.~S.}\
  \bibnamefont {Holevo}},\ and\ \bibinfo {editor} {\bibfnamefont {C.~M.}\
  \bibnamefont {Caves}}}\ (\bibinfo  {publisher} {Springer US},\ \bibinfo
  {address} {Boston, MA},\ \bibinfo {year} {1997})\ pp.\ \bibinfo {pages}
  {181--188}\BibitemShut {NoStop}%
\bibitem [{\citenamefont {Preskill}(2018)}]{Preskill2018NISQ}%
  \BibitemOpen
  \bibfield  {author} {\bibinfo {author} {\bibfnamefont {J.}~\bibnamefont
  {Preskill}},\ }\href {https://doi.org/10.22331/q-2018-08-06-79} {\bibfield
  {journal} {\bibinfo  {journal} {Quantum}\ }\textbf {\bibinfo {volume} {2}},\
  \bibinfo {pages} {79} (\bibinfo {year} {2018})}\BibitemShut {NoStop}%
\bibitem [{\citenamefont {Bultrini}\ \emph {et~al.}(2022)\citenamefont
  {Bultrini}, \citenamefont {Wang}, \citenamefont {Czarnik}, \citenamefont
  {Gordon}, \citenamefont {Cerezo}, \citenamefont {Coles},\ and\ \citenamefont
  {Cincio}}]{Bultrini2022}%
  \BibitemOpen
  \bibfield  {author} {\bibinfo {author} {\bibfnamefont {D.}~\bibnamefont
  {Bultrini}}, \bibinfo {author} {\bibfnamefont {S.}~\bibnamefont {Wang}},
  \bibinfo {author} {\bibfnamefont {P.}~\bibnamefont {Czarnik}}, \bibinfo
  {author} {\bibfnamefont {M.~H.}\ \bibnamefont {Gordon}}, \bibinfo {author}
  {\bibfnamefont {M.}~\bibnamefont {Cerezo}}, \bibinfo {author} {\bibfnamefont
  {P.~J.}\ \bibnamefont {Coles}},\ and\ \bibinfo {author} {\bibfnamefont
  {L.}~\bibnamefont {Cincio}},\ }\href {https://arxiv.org/abs/0707.1883} {\
  (\bibinfo {year} {2022})},\ \Eprint {https://arxiv.org/abs/arXiv:2205.13454}
  {arXiv:arXiv:2205.13454 [quant-ph]} \BibitemShut {NoStop}%
\bibitem [{\citenamefont {Aharonov}\ and\ \citenamefont
  {Ben-Or}(1996)}]{aharonov1996}%
  \BibitemOpen
  \bibfield  {author} {\bibinfo {author} {\bibfnamefont {D.}~\bibnamefont
  {Aharonov}}\ and\ \bibinfo {author} {\bibfnamefont {M.}~\bibnamefont
  {Ben-Or}},\ }\href {https://arxiv.org/abs/quant-ph/9611025} {\bibinfo {title}
  {Fault tolerant quantum computation with constant error}} (\bibinfo {year}
  {1996}),\ \Eprint {https://arxiv.org/abs/arXiv:quant-ph/9611025}
  {arXiv:arXiv:quant-ph/9611025 [quant-ph]} \BibitemShut {NoStop}%
\bibitem [{\citenamefont {Knill}\ \emph {et~al.}(1996)\citenamefont {Knill},
  \citenamefont {Laflamme},\ and\ \citenamefont {Zurek}}]{knill1996}%
  \BibitemOpen
  \bibfield  {author} {\bibinfo {author} {\bibfnamefont {E.}~\bibnamefont
  {Knill}}, \bibinfo {author} {\bibfnamefont {R.}~\bibnamefont {Laflamme}},\
  and\ \bibinfo {author} {\bibfnamefont {W.}~\bibnamefont {Zurek}},\ }\href
  {https://arxiv.org/abs/quant-ph/9610011} {\bibinfo {title} {Threshold
  accuracy for quantum computation}} (\bibinfo {year} {1996}),\ \Eprint
  {https://arxiv.org/abs/arXiv:quant-ph/9610011} {arXiv:arXiv:quant-ph/9610011
  [quant-ph]} \BibitemShut {NoStop}%
\bibitem [{\citenamefont {Lamata}\ \emph {et~al.}(2018)\citenamefont {Lamata},
  \citenamefont {Parra-Rodriguez}, \citenamefont {Sanz},\ and\ \citenamefont
  {Solano}}]{Lamata2018}%
  \BibitemOpen
  \bibfield  {author} {\bibinfo {author} {\bibfnamefont {L.}~\bibnamefont
  {Lamata}}, \bibinfo {author} {\bibfnamefont {A.}~\bibnamefont
  {Parra-Rodriguez}}, \bibinfo {author} {\bibfnamefont {M.}~\bibnamefont
  {Sanz}},\ and\ \bibinfo {author} {\bibfnamefont {E.}~\bibnamefont {Solano}},\
  }\href {https://doi.org/10.1080/23746149.2018.1457981} {\bibfield  {journal}
  {\bibinfo  {journal} {Adv. Phys. X}\ }\textbf {\bibinfo {volume} {3}},\
  \bibinfo {pages} {1457981} (\bibinfo {year} {2018})}\BibitemShut {NoStop}%
\bibitem [{\citenamefont {Parra-Rodriguez}\ \emph {et~al.}(2020)\citenamefont
  {Parra-Rodriguez}, \citenamefont {Lougovski}, \citenamefont {Lamata},
  \citenamefont {Solano},\ and\ \citenamefont {Sanz}}]{Adrian2020DAQC}%
  \BibitemOpen
  \bibfield  {author} {\bibinfo {author} {\bibfnamefont {A.}~\bibnamefont
  {Parra-Rodriguez}}, \bibinfo {author} {\bibfnamefont {P.}~\bibnamefont
  {Lougovski}}, \bibinfo {author} {\bibfnamefont {L.}~\bibnamefont {Lamata}},
  \bibinfo {author} {\bibfnamefont {E.}~\bibnamefont {Solano}},\ and\ \bibinfo
  {author} {\bibfnamefont {M.}~\bibnamefont {Sanz}},\ }\href
  {https://doi.org/10.1103/physreva.101.022305} {\bibfield  {journal} {\bibinfo
   {journal} {Phys. Rev. A}\ }\textbf {\bibinfo {volume} {101}},\ \bibinfo
  {pages} {022305} (\bibinfo {year} {2020})}\BibitemShut {NoStop}%
\bibitem [{\citenamefont {García-Molina}\ \emph {et~al.}(2022)\citenamefont
  {García-Molina}, \citenamefont {Martin}, \citenamefont {Garcia-de Andoin},\
  and\ \citenamefont {Sanz}}]{garcia2022noise}%
  \BibitemOpen
  \bibfield  {author} {\bibinfo {author} {\bibfnamefont {P.}~\bibnamefont
  {García-Molina}}, \bibinfo {author} {\bibfnamefont {A.}~\bibnamefont
  {Martin}}, \bibinfo {author} {\bibfnamefont {M.}~\bibnamefont {Garcia-de
  Andoin}},\ and\ \bibinfo {author} {\bibfnamefont {M.}~\bibnamefont {Sanz}},\
  }\href {https://arxiv.org/abs/2107.12969} {\  (\bibinfo {year} {2022})},\
  \Eprint {https://arxiv.org/abs/arXiv:2107.12969} {arXiv:arXiv:2107.12969
  [quant-ph]} \BibitemShut {NoStop}%
\bibitem [{\citenamefont {Gong}\ \emph {et~al.}(2023)\citenamefont {Gong},
  \citenamefont {Huang}, \citenamefont {Wang}, \citenamefont {Guo},
  \citenamefont {Li}, \citenamefont {Wu}, \citenamefont {Zhu}, \citenamefont
  {Zhao}, \citenamefont {Guo}, \citenamefont {Qian}, \citenamefont {Ye},
  \citenamefont {Zha}, \citenamefont {Chen}, \citenamefont {Ying},
  \citenamefont {Yu}, \citenamefont {Fan}, \citenamefont {Wu}, \citenamefont
  {Su}, \citenamefont {Deng}, \citenamefont {Rong}, \citenamefont {Zhang},
  \citenamefont {Cao}, \citenamefont {Lin}, \citenamefont {Xu}, \citenamefont
  {Sun}, \citenamefont {Guo}, \citenamefont {Li}, \citenamefont {Liang},
  \citenamefont {Sakurai}, \citenamefont {Nemoto}, \citenamefont {Munro},
  \citenamefont {Huo}, \citenamefont {Lu}, \citenamefont {Peng}, \citenamefont
  {Zhu},\ and\ \citenamefont {Pan}}]{WeiPan202261qubitDAQC}%
  \BibitemOpen
  \bibfield  {author} {\bibinfo {author} {\bibfnamefont {M.}~\bibnamefont
  {Gong}}, \bibinfo {author} {\bibfnamefont {H.-L.}\ \bibnamefont {Huang}},
  \bibinfo {author} {\bibfnamefont {S.}~\bibnamefont {Wang}}, \bibinfo {author}
  {\bibfnamefont {C.}~\bibnamefont {Guo}}, \bibinfo {author} {\bibfnamefont
  {S.}~\bibnamefont {Li}}, \bibinfo {author} {\bibfnamefont {Y.}~\bibnamefont
  {Wu}}, \bibinfo {author} {\bibfnamefont {Q.}~\bibnamefont {Zhu}}, \bibinfo
  {author} {\bibfnamefont {Y.}~\bibnamefont {Zhao}}, \bibinfo {author}
  {\bibfnamefont {S.}~\bibnamefont {Guo}}, \bibinfo {author} {\bibfnamefont
  {H.}~\bibnamefont {Qian}}, \bibinfo {author} {\bibfnamefont {Y.}~\bibnamefont
  {Ye}}, \bibinfo {author} {\bibfnamefont {C.}~\bibnamefont {Zha}}, \bibinfo
  {author} {\bibfnamefont {F.}~\bibnamefont {Chen}}, \bibinfo {author}
  {\bibfnamefont {C.}~\bibnamefont {Ying}}, \bibinfo {author} {\bibfnamefont
  {J.}~\bibnamefont {Yu}}, \bibinfo {author} {\bibfnamefont {D.}~\bibnamefont
  {Fan}}, \bibinfo {author} {\bibfnamefont {D.}~\bibnamefont {Wu}}, \bibinfo
  {author} {\bibfnamefont {H.}~\bibnamefont {Su}}, \bibinfo {author}
  {\bibfnamefont {H.}~\bibnamefont {Deng}}, \bibinfo {author} {\bibfnamefont
  {H.}~\bibnamefont {Rong}}, \bibinfo {author} {\bibfnamefont {K.}~\bibnamefont
  {Zhang}}, \bibinfo {author} {\bibfnamefont {S.}~\bibnamefont {Cao}}, \bibinfo
  {author} {\bibfnamefont {J.}~\bibnamefont {Lin}}, \bibinfo {author}
  {\bibfnamefont {Y.}~\bibnamefont {Xu}}, \bibinfo {author} {\bibfnamefont
  {L.}~\bibnamefont {Sun}}, \bibinfo {author} {\bibfnamefont {C.}~\bibnamefont
  {Guo}}, \bibinfo {author} {\bibfnamefont {N.}~\bibnamefont {Li}}, \bibinfo
  {author} {\bibfnamefont {F.}~\bibnamefont {Liang}}, \bibinfo {author}
  {\bibfnamefont {A.}~\bibnamefont {Sakurai}}, \bibinfo {author} {\bibfnamefont
  {K.}~\bibnamefont {Nemoto}}, \bibinfo {author} {\bibfnamefont {W.~J.}\
  \bibnamefont {Munro}}, \bibinfo {author} {\bibfnamefont {Y.-H.}\ \bibnamefont
  {Huo}}, \bibinfo {author} {\bibfnamefont {C.-Y.}\ \bibnamefont {Lu}},
  \bibinfo {author} {\bibfnamefont {C.-Z.}\ \bibnamefont {Peng}}, \bibinfo
  {author} {\bibfnamefont {X.}~\bibnamefont {Zhu}},\ and\ \bibinfo {author}
  {\bibfnamefont {J.-W.}\ \bibnamefont {Pan}},\ }\href
  {https://doi.org/https://doi.org/10.1016/j.scib.2023.04.003} {\bibfield
  {journal} {\bibinfo  {journal} {Sci. Bull.}\ }\textbf {\bibinfo {volume}
  {68}},\ \bibinfo {pages} {906} (\bibinfo {year} {2023})}\BibitemShut
  {NoStop}%
\bibitem [{\citenamefont {Gonzalez-Raya}\ \emph {et~al.}(2021)\citenamefont
  {Gonzalez-Raya}, \citenamefont {Asensio-Perea}, \citenamefont {Martin},
  \citenamefont {C\'eleri}, \citenamefont {Sanz}, \citenamefont {Lougovski},\
  and\ \citenamefont {Dumitrescu}}]{Tasio2021DAQCcrossResonance}%
  \BibitemOpen
  \bibfield  {author} {\bibinfo {author} {\bibfnamefont {T.}~\bibnamefont
  {Gonzalez-Raya}}, \bibinfo {author} {\bibfnamefont {R.}~\bibnamefont
  {Asensio-Perea}}, \bibinfo {author} {\bibfnamefont {A.}~\bibnamefont
  {Martin}}, \bibinfo {author} {\bibfnamefont {L.~C.}\ \bibnamefont
  {C\'eleri}}, \bibinfo {author} {\bibfnamefont {M.}~\bibnamefont {Sanz}},
  \bibinfo {author} {\bibfnamefont {P.}~\bibnamefont {Lougovski}},\ and\
  \bibinfo {author} {\bibfnamefont {E.~F.}\ \bibnamefont {Dumitrescu}},\ }\href
  {https://doi.org/10.1103/PRXQuantum.2.020328} {\bibfield  {journal} {\bibinfo
   {journal} {PRX Quantum}\ }\textbf {\bibinfo {volume} {2}},\ \bibinfo {pages}
  {020328} (\bibinfo {year} {2021})}\BibitemShut {NoStop}%
\bibitem [{\citenamefont {Galicia}\ \emph {et~al.}(2020)\citenamefont
  {Galicia}, \citenamefont {Ramon}, \citenamefont {Solano},\ and\ \citenamefont
  {Sanz}}]{Galicia2020EnhancedConnect}%
  \BibitemOpen
  \bibfield  {author} {\bibinfo {author} {\bibfnamefont {A.}~\bibnamefont
  {Galicia}}, \bibinfo {author} {\bibfnamefont {B.}~\bibnamefont {Ramon}},
  \bibinfo {author} {\bibfnamefont {E.}~\bibnamefont {Solano}},\ and\ \bibinfo
  {author} {\bibfnamefont {M.}~\bibnamefont {Sanz}},\ }\href
  {https://doi.org/10.1103/PhysRevResearch.2.033103} {\bibfield  {journal}
  {\bibinfo  {journal} {Phys. Rev. Research}\ }\textbf {\bibinfo {volume}
  {2}},\ \bibinfo {pages} {033103} (\bibinfo {year} {2020})}\BibitemShut
  {NoStop}%
\bibitem [{\citenamefont {Dodd}\ \emph {et~al.}(2002)\citenamefont {Dodd},
  \citenamefont {Nielsen}, \citenamefont {Bremner},\ and\ \citenamefont
  {Thew}}]{Dodd2002UnivQC}%
  \BibitemOpen
  \bibfield  {author} {\bibinfo {author} {\bibfnamefont {J.~L.}\ \bibnamefont
  {Dodd}}, \bibinfo {author} {\bibfnamefont {M.~A.}\ \bibnamefont {Nielsen}},
  \bibinfo {author} {\bibfnamefont {M.~J.}\ \bibnamefont {Bremner}},\ and\
  \bibinfo {author} {\bibfnamefont {R.~T.}\ \bibnamefont {Thew}},\ }\href
  {https://doi.org/10.1103/PhysRevA.65.040301} {\bibfield  {journal} {\bibinfo
  {journal} {Phys. Rev. A}\ }\textbf {\bibinfo {volume} {65}},\ \bibinfo
  {pages} {040301} (\bibinfo {year} {2002})}\BibitemShut {NoStop}%
\bibitem [{\citenamefont {Suzuki}(1976)}]{Suzuki1976Trotter}%
  \BibitemOpen
  \bibfield  {author} {\bibinfo {author} {\bibfnamefont {M.}~\bibnamefont
  {Suzuki}},\ }\href {https://doi.org/10.1007/BF01609348} {\bibfield  {journal}
  {\bibinfo  {journal} {Commun. Math. Phys.}\ }\textbf {\bibinfo {volume}
  {51}},\ \bibinfo {pages} {183} (\bibinfo {year} {1976})}\BibitemShut
  {NoStop}%
\bibitem [{\citenamefont {Ba{\ss{}}ler}\ \emph {et~al.}(2023)\citenamefont
  {Ba{\ss{}}ler}, \citenamefont {Zipper}, \citenamefont {Cedzich},
  \citenamefont {Heinrich}, \citenamefont {Huber}, \citenamefont {Johanning},\
  and\ \citenamefont {Kliesch}}]{Pascal2023}%
  \BibitemOpen
  \bibfield  {author} {\bibinfo {author} {\bibfnamefont {P.}~\bibnamefont
  {Ba{\ss{}}ler}}, \bibinfo {author} {\bibfnamefont {M.}~\bibnamefont
  {Zipper}}, \bibinfo {author} {\bibfnamefont {C.}~\bibnamefont {Cedzich}},
  \bibinfo {author} {\bibfnamefont {M.}~\bibnamefont {Heinrich}}, \bibinfo
  {author} {\bibfnamefont {P.~H.}\ \bibnamefont {Huber}}, \bibinfo {author}
  {\bibfnamefont {M.}~\bibnamefont {Johanning}},\ and\ \bibinfo {author}
  {\bibfnamefont {M.}~\bibnamefont {Kliesch}},\ }\href
  {https://doi.org/10.22331/q-2023-04-20-984} {\bibfield  {journal} {\bibinfo
  {journal} {{Quantum}}\ }\textbf {\bibinfo {volume} {7}},\ \bibinfo {pages}
  {984} (\bibinfo {year} {2023})}\BibitemShut {NoStop}%
\bibitem [{\citenamefont {Lawson}\ and\ \citenamefont
  {Hanson}(1974{\natexlab{a}})}]{Lawson1974}%
  \BibitemOpen
  \bibfield  {author} {\bibinfo {author} {\bibfnamefont {C.~L.}\ \bibnamefont
  {Lawson}}\ and\ \bibinfo {author} {\bibfnamefont {R.~J.}\ \bibnamefont
  {Hanson}},\ }\bibinfo {title} {Solving least squares problems}\ (\bibinfo
  {publisher} {Pentice-Hall},\ \bibinfo {year} {1974})\ Chap.~\bibinfo
  {chapter} {23}, pp.\ \bibinfo {pages} {158--173}\BibitemShut {NoStop}%
\bibitem [{\citenamefont {Vidal}(2003)}]{Vidal2003}%
  \BibitemOpen
  \bibfield  {author} {\bibinfo {author} {\bibfnamefont {G.}~\bibnamefont
  {Vidal}},\ }\href {https://doi.org/10.1103/physrevlett.91.147902} {\bibfield
  {journal} {\bibinfo  {journal} {Phys. Rev. Lett.}\ }\textbf {\bibinfo
  {volume} {91}},\ \bibinfo {pages} {147902} (\bibinfo {year}
  {2003})}\BibitemShut {NoStop}%
\bibitem [{\citenamefont {García-Molina}\ \emph {et~al.}(2023)\citenamefont
  {García-Molina}, \citenamefont {Tagliacozzo},\ and\ \citenamefont
  {García-Ripoll}}]{GarciaMolina2023}%
  \BibitemOpen
  \bibfield  {author} {\bibinfo {author} {\bibfnamefont {P.}~\bibnamefont
  {García-Molina}}, \bibinfo {author} {\bibfnamefont {L.}~\bibnamefont
  {Tagliacozzo}},\ and\ \bibinfo {author} {\bibfnamefont {J.~J.}\ \bibnamefont
  {García-Ripoll}},\ }\href {https://arxiv.org/abs/2303.09430} {\  (\bibinfo
  {year} {2023})},\ \Eprint {https://arxiv.org/abs/arXiv:2303.09430}
  {arXiv:arXiv:2303.09430 [quant-ph]} \BibitemShut {NoStop}%
\bibitem [{\citenamefont {Hegade}\ \emph {et~al.}(2021)\citenamefont {Hegade},
  \citenamefont {Paul}, \citenamefont {Ding}, \citenamefont {Sanz},
  \citenamefont {Albarr\'an-Arriagada}, \citenamefont {Solano},\ and\
  \citenamefont {Chen}}]{STADAQC}%
  \BibitemOpen
  \bibfield  {author} {\bibinfo {author} {\bibfnamefont {N.~N.}\ \bibnamefont
  {Hegade}}, \bibinfo {author} {\bibfnamefont {K.}~\bibnamefont {Paul}},
  \bibinfo {author} {\bibfnamefont {Y.}~\bibnamefont {Ding}}, \bibinfo {author}
  {\bibfnamefont {M.}~\bibnamefont {Sanz}}, \bibinfo {author} {\bibfnamefont
  {F.}~\bibnamefont {Albarr\'an-Arriagada}}, \bibinfo {author} {\bibfnamefont
  {E.}~\bibnamefont {Solano}},\ and\ \bibinfo {author} {\bibfnamefont
  {X.}~\bibnamefont {Chen}},\ }\href
  {https://doi.org/10.1103/PhysRevApplied.15.024038} {\bibfield  {journal}
  {\bibinfo  {journal} {Phys. Rev. Appl.}\ }\textbf {\bibinfo {volume} {15}},\
  \bibinfo {pages} {024038} (\bibinfo {year} {2021})}\BibitemShut {NoStop}%
\bibitem [{\citenamefont {Altenberg}(1995)}]{GenAlg}%
  \BibitemOpen
  \bibfield  {author} {\bibinfo {author} {\bibfnamefont {L.}~\bibnamefont
  {Altenberg}},\ }\bibinfo {title} {The {S}chema theorem and {P}rice's
  theorem}\ (\bibinfo  {publisher} {Elsevier},\ \bibinfo {year} {1995})\ pp.\
  \bibinfo {pages} {23--49}\BibitemShut {NoStop}%
\bibitem [{\citenamefont {Kennedy}\ and\ \citenamefont
  {Eberhart}(1995)}]{SwarmAlg}%
  \BibitemOpen
  \bibfield  {author} {\bibinfo {author} {\bibfnamefont {J.}~\bibnamefont
  {Kennedy}}\ and\ \bibinfo {author} {\bibfnamefont {R.}~\bibnamefont
  {Eberhart}},\ }in\ \href {https://doi.org/10.1109/ICNN.1995.488968} {\emph
  {\bibinfo {booktitle} {Proc. ICNN'95}}},\ Vol.~\bibinfo {volume} {4}\
  (\bibinfo {year} {1995})\ pp.\ \bibinfo {pages} {1942--1948
  vol.4}\BibitemShut {NoStop}%
\bibitem [{\citenamefont {Mockus}(1989)}]{Bayesian_Mockus_1989}%
  \BibitemOpen
  \bibfield  {author} {\bibinfo {author} {\bibfnamefont {J.}~\bibnamefont
  {Mockus}},\ }\href {https://doi.org/10.1007/978-94-009-0909-0} {\emph
  {\bibinfo {title} {Bayesian Approach to Global Optimization}}}\ (\bibinfo
  {publisher} {Springer Netherlands},\ \bibinfo {year} {1989})\BibitemShut
  {NoStop}%
\bibitem [{\citenamefont {Frazier}(2018)}]{Bayesian_guide}%
  \BibitemOpen
  \bibfield  {author} {\bibinfo {author} {\bibfnamefont {P.~I.}\ \bibnamefont
  {Frazier}},\ }\href {https://arxiv.org/abs/1807.02811} {\bibinfo {title} {A
  tutorial on {B}ayesian optimization}} (\bibinfo {year} {2018}),\ \Eprint
  {https://arxiv.org/abs/arXiv:1807.02811} {arXiv:arXiv:1807.02811 [quant-ph]}
  \BibitemShut {NoStop}%
\bibitem [{\citenamefont {Feurer}\ and\ \citenamefont {Hutter}(2019)}]{hypop}%
  \BibitemOpen
  \bibfield  {author} {\bibinfo {author} {\bibfnamefont {M.}~\bibnamefont
  {Feurer}}\ and\ \bibinfo {author} {\bibfnamefont {F.}~\bibnamefont
  {Hutter}},\ }\bibinfo {title} {Hyperparameter optimization}\ (\bibinfo
  {publisher} {Springer International Publishing},\ \bibinfo {year} {2019})\
  pp.\ \bibinfo {pages} {3--33}\BibitemShut {NoStop}%
\bibitem [{Discovered thanks to the OEIS()}]{APX:OEIS}%
  \BibitemOpen
  Discovered thanks to the OEIS,\ \href {https://oeis.org/A002024} {}\bibinfo
  {howpublished} {https://oeis.org/A002024}\BibitemShut {NoStop}%
\bibitem [{Note1()}]{Note1}%
  \BibitemOpen
  \bibinfo {note} {The formal determinant has exactly the same expression as a
  usual determinant, but instead of the elements being numbers, they are square
  matrices of equal size that commute with each other by pairs.}\BibitemShut
  {Stop}%
\bibitem [{\citenamefont {Ingraham}(1937)}]{APX:Ingraham1937}%
  \BibitemOpen
  \bibfield  {author} {\bibinfo {author} {\bibfnamefont {M.~H.}\ \bibnamefont
  {Ingraham}},\ }\href {https://doi.org/10.1090/S0002-9904-1937-06609-2}
  {\bibfield  {journal} {\bibinfo  {journal} {Bull. Amer. Math. Soc.}\ }\textbf
  {\bibinfo {volume} {43}},\ \bibinfo {pages} {579} (\bibinfo {year}
  {1937})}\BibitemShut {NoStop}%
\bibitem [{Note2()}]{Note2}%
  \BibitemOpen
  \bibinfo {note} {$(a+b)\protect \mod x=(a\protect \mod x+b\protect \mod
  x)\protect \mod x$}\BibitemShut {NoStop}%
\bibitem [{Note3()}]{Note3}%
  \BibitemOpen
  \bibinfo {note} {$ab\protect \mod x=[(a\protect \mod x)(b\protect \mod
  x)]\protect \mod x$}\BibitemShut {NoStop}%
\bibitem [{\citenamefont {Lawson}\ and\ \citenamefont
  {Hanson}(1974{\natexlab{b}})}]{APX:Lawson1974}%
  \BibitemOpen
  \bibfield  {author} {\bibinfo {author} {\bibfnamefont {C.~L.}\ \bibnamefont
  {Lawson}}\ and\ \bibinfo {author} {\bibfnamefont {R.~J.}\ \bibnamefont
  {Hanson}},\ }\bibinfo {title} {Solving least-squares problems}\ (\bibinfo
  {publisher} {Prentice Hall},\ \bibinfo {year} {1974})\ Chap.~\bibinfo
  {chapter} {23}\BibitemShut {NoStop}%
\bibitem [{\citenamefont {Dobbins}(2015)}]{APX:Dobbins2015}%
  \BibitemOpen
  \bibfield  {author} {\bibinfo {author} {\bibfnamefont {M.~G.}\ \bibnamefont
  {Dobbins}},\ }\href {https://doi.org/10.1007/s00222-014-0523-2} {\bibfield
  {journal} {\bibinfo  {journal} {Invent. Math.}\ }\textbf {\bibinfo {volume}
  {199}},\ \bibinfo {pages} {287} (\bibinfo {year} {2015})}\BibitemShut
  {NoStop}%
\bibitem [{\citenamefont {Blagojević}\ \emph {et~al.}(2019)\citenamefont
  {Blagojević}, \citenamefont {Frick},\ and\ \citenamefont
  {Ziegler}}]{APX:Blagojevic2015}%
  \BibitemOpen
  \bibfield  {author} {\bibinfo {author} {\bibfnamefont {P.~V.~M.}\
  \bibnamefont {Blagojević}}, \bibinfo {author} {\bibfnamefont
  {F.}~\bibnamefont {Frick}},\ and\ \bibinfo {author} {\bibfnamefont {G.~M.}\
  \bibnamefont {Ziegler}},\ }\href {https://doi.org/10.4171/jems/881}
  {\bibfield  {journal} {\bibinfo  {journal} {J. Eur. Math. Soc.}\ }\textbf
  {\bibinfo {volume} {21}},\ \bibinfo {pages} {2107} (\bibinfo {year}
  {2019})}\BibitemShut {NoStop}%
\end{thebibliography}%

\end{document}